\documentclass[runningheads]{llncs}

\usepackage[year=2026]{eccv}
\usepackage{eccvabbrv}
\usepackage[leftcaption,raggedright]{sidecap}
\sidecaptionvpos{figure}{t}
\usepackage[font=small,labelfont=bf]{caption}
\usepackage{graphicx}

\usepackage{amsmath}

\usepackage{amssymb}

\usepackage{tabularx}
\usepackage{booktabs,array}
\usepackage[table]{xcolor}
\usepackage{colortbl}
\usepackage{adjustbox}

\definecolor{tblHead}{gray}{0.92}
\definecolor{tblSect}{gray}{0.95}
\definecolor{tblLite}{gray}{0.975} 
\definecolor{seedHL}{RGB}{255,248,220}

\newcolumntype{L}[1]{>{\raggedright\arraybackslash}p{#1}}
\newcolumntype{Y}{>{\raggedright\arraybackslash}X}

\newcommand{\SectRowS}[1]{%
\rowcolor{tblSect}\multicolumn{3}{@{}l}{\rule{0pt}{1.05em}\textbf{#1}}\\[-0.25em]}

\newcommand{\beginsupp}{%
  \setcounter{table}{0}%
  \setcounter{figure}{0}%
  \renewcommand{\thetable}{S\arabic{table}}%
  \renewcommand{\thefigure}{S\arabic{figure}}%
}

\usepackage{booktabs,array,makecell,xcolor,colortbl,pifont}

\definecolor{tblHead}{gray}{0.92}
\definecolor{tblSect}{gray}{0.95}
\definecolor{seedHL}{RGB}{255,248,220} 

\newcolumntype{L}[1]{>{\raggedright\arraybackslash}p{#1}}
\newcolumntype{C}[1]{>{\centering\arraybackslash}p{#1}}

\newcommand{\yes}{\ding{51}} 

\newcommand{\SectRow}[1]{%
\rowcolor{tblSect}\multicolumn{11}{@{}l}{\rule{0pt}{1.05em}\textbf{#1}}\\[-0.25em]}

\usepackage{microtype}

\usepackage{multirow}
\usepackage{makecell}     
\usepackage{adjustbox}

\usepackage{subcaption}
\usepackage{stfloats}
\usepackage{enumitem}

\newcolumntype{P}[1]{>{\raggedright\arraybackslash}p{#1}}
\newcolumntype{C}[1]{>{\centering\arraybackslash}p{#1}}
\newcolumntype{Y}{>{\raggedright\arraybackslash}X}

\usepackage{marvosym}
\usepackage{url}

\usepackage[accsupp]{axessibility}  

\usepackage{orcidlink}

\newcolumntype{L}[1]{>{\raggedright\arraybackslash\hspace{0pt}}p{#1}}
\newcolumntype{C}[1]{>{\centering\arraybackslash}p{#1}}

\definecolor{faithblue}{RGB}{230,245,255}
\newcommand{\best}[1]{\textbf{#1}}
\newcommand{\second}[1]{\underline{#1}}

\begin{document}

\title{\raisebox{-0.25em}{\includegraphics[height=1.2em]{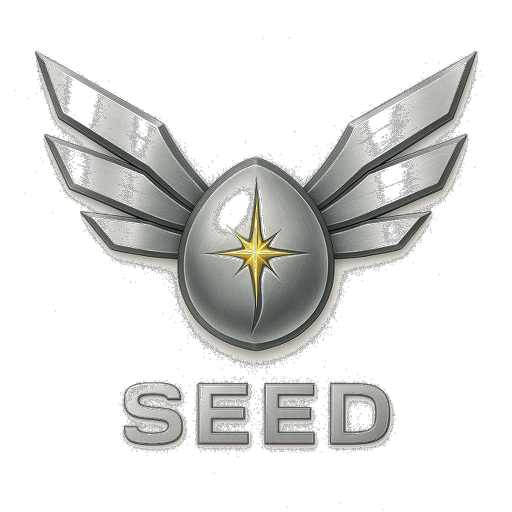}}SEED: A Large-Scale Benchmark for Provenance Tracing in Sequential Deepfake Facial Edits}

\author{
Mengieong Hoi\inst{1}
\and
Zhedong Zheng\inst{1}
\and
Ping Liu\inst{2}\Letter
\and
Wei Liu\inst{3}\Letter
}

\institute{
University of Macau, Macau, China\\
\email{mc45294@um.edu.mo, zhedongzheng@um.edu.mo}
\and
University of Nevada, Reno, NV, USA\\
\email{pino.pingliu@gmail.com}
\and
Huazhong University of Science and Technology, Wuhan, China\\
\email{liuwei@hust.edu.cn}\\[0.3em]
\Letter\ Corresponding authors
}

\maketitle

\begin{abstract}
Deepfake content on social networks is increasingly produced through multiple \emph{sequential} edits to biometric data such as facial imagery. Consequently, the final appearance of an image often reflects a latent chain of operations rather than a single manipulation. Recovering these editing histories is essential for visual provenance analysis, misinformation auditing, and forensic or platform moderation workflows that must trace the origin and evolution of AI-generated media. However, existing datasets predominantly focus on single-step editing and overlook the cumulative artifacts introduced by realistic multi-step pipelines. To address this gap, we introduce Sequential Editing in Diffusion (\textbf{SEED}), a large-scale benchmark for sequential provenance tracing in facial imagery. SEED contains over 90K images constructed via one to four sequential attribute edits using diffusion-based editing pipelines, with fine-grained annotations including edit order, textual instructions, manipulation masks, and generation models. These metadata enable step-wise evidence analysis and support forgery detection, sequence prediction. To benchmark the challenges posed by SEED, we evaluate representative analysis strategies and observe that spatial-only approaches struggle under subtle and distributed diffusion artifacts, especially when such artifacts accumulate across multiple edits. Motivated by this observation, we further establish \textbf{FAITH}, a frequency-aware Transformer baseline that aggregates spatial and frequency-domain cues to identify and order latent editing events. Results show that high-frequency signals, particularly wavelet components, provide effective cues even under image degradation. Overall, SEED facilitates systematic study of sequential provenance tracing and evidence aggregation for trustworthy analysis of AI-generated visual content.
\end{abstract}

\begin{figure}[t]
    \centering
    \vspace{-.1in}
    \includegraphics[width=0.9\linewidth]{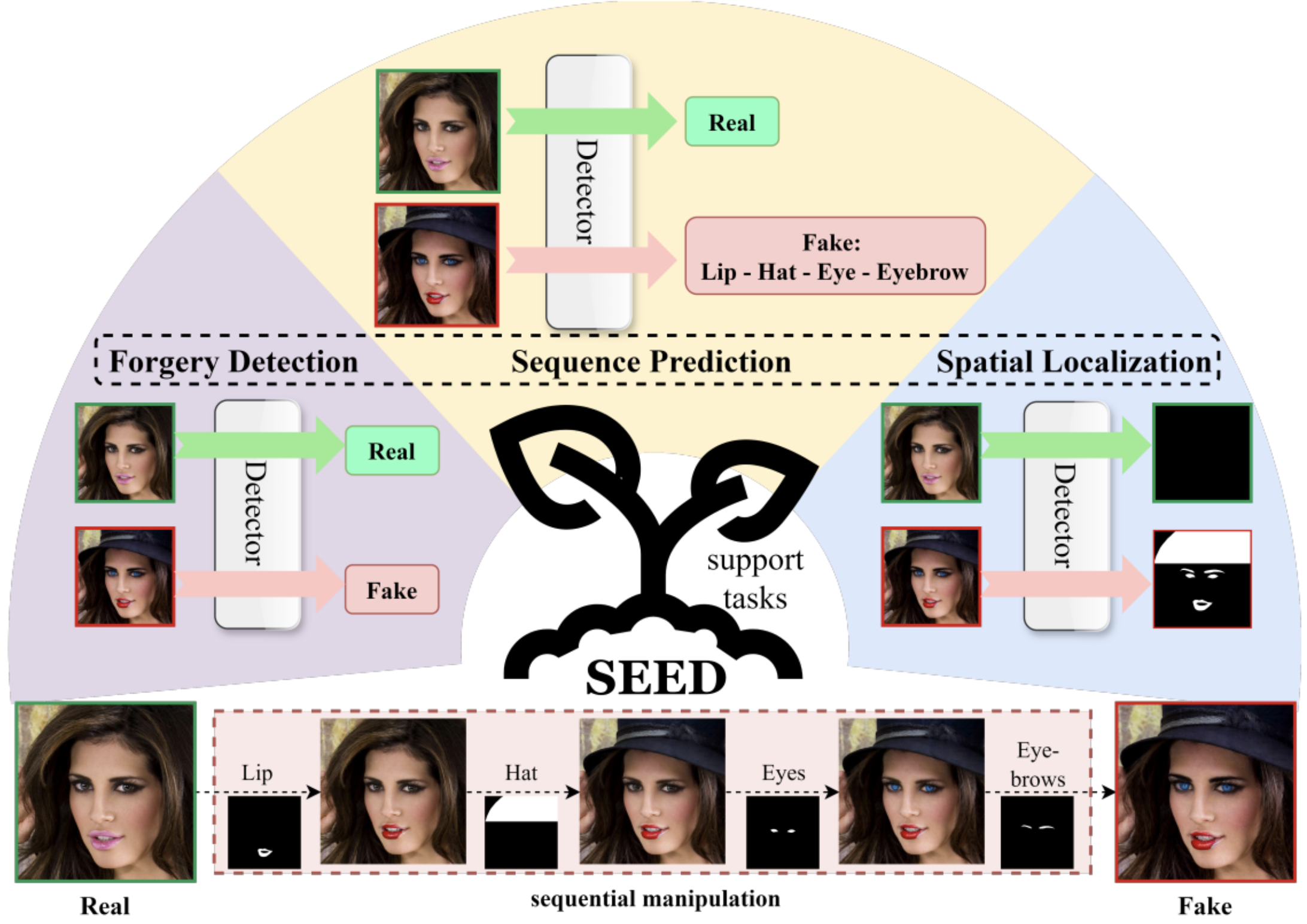}
    \vspace{-.1in}
    \caption{Overview of SEED and supported provenance analysis tasks.
SEED targets diffusion-based \emph{sequential} facial editing and provides supervision for three complementary tasks:
\emph{Authenticity Analysis}, which distinguishes real images from sequentially edited ones;
\emph{Editing Trace Analysis}, which characterizes the edited attributes and their sequential patterns;
and \emph{Spatial Evidence Analysis}, which analyzes manipulated regions using edit masks.
The bottom example illustrates a real face progressively edited into a fake through a sequence of attribute edits such as lip, hat, eyes, and eyebrows, together with corresponding manipulation masks that support fine-grained evidence analysis.}
    \label{fig:seed_overview}
    \vspace{-5mm}
\end{figure}

\begin{table}[t]
\scriptsize
\centering
\caption{Comparison of image-based deepfake/AIGC datasets by supervision and supported tasks. Scale is reported in thousands of images (K) without separating real/fake.}
\vspace{-.1in}
\label{tab:forgery_datasets_tick}
\setlength{\tabcolsep}{1.9pt}
\renewcommand{\arraystretch}{1.10}
\begingroup
\hyphenpenalty=10000
\exhyphenpenalty=10000
\sloppy

\begin{tabular}{@{}
L{3.5cm} C{0.95cm} C{0.95cm} C{0.65cm}
*{4}{C{0.6cm}} *{3}{C{0.6cm}}
}
\toprule
\rowcolor{tblHead}
\textbf{Dataset} &
\textbf{Mod.} &
\textbf{Scale} &
\textbf{Seq} &
\multicolumn{4}{c}{\textbf{Supervision}} &
\multicolumn{3}{c}{\textbf{Tasks}} \\
\rowcolor{tblHead}
& & & &
\scriptsize Ord & \scriptsize Attr & \scriptsize Txt & \scriptsize Loc &
\scriptsize Det & \scriptsize Prov & \scriptsize Loc \\
\cmidrule(lr){5-8}\cmidrule(lr){9-11}

\SectRow{A. Image Detection}
OpenForensics~\cite{le2021openforensics} &
I &
115.8K &
- &
& & & \yes &
\yes & & \yes \\

ForgeryNet~\cite{he2021forgerynet} &
I &
2900.0K &
- &
& & & \yes &
\yes & & \yes \\

DeepFakeFace-DFF~\cite{song2023deepfakeface} &
I &
120.0K &
- &
& & & &
\yes & & \\

DiffusionFace~\cite{chen2024diffusionface} &
I &
630.0K &
- &
& & \yes & \yes &
\yes & & \yes \\

GenImage~\cite{zhu2023genimage} &
I &
2000.0K$^{\dagger}$ &
- &
& & & &
\yes & & \\

DRCT-2M~\cite{chen2024drct} &
I &
2000.0K &
- &
& & & &
\yes & & \\

ImagiNet~\cite{boychev2024imaginet} &
I &
200.0K &
- &
& & & &
\yes & & \\

Fake2M~\cite{lu2023seeing} &
I &
2087.0K &
- &
& & \yes & &
\yes & & \\

\addlinespace[0.15em]
\midrule
\SectRow{B. Provenance Tracing}
Seq-DeepFake-Comp~\cite{shao2022sequential} &
I &
35.2K &
28 &
\yes & \yes & & &
\yes & \yes & \\

Seq-DeepFake-Attr~\cite{shao2022sequential} &
I &
49.9K &
26 &
\yes & \yes & & &
\yes & \yes & \\

Seq-DeepFake-P~\cite{shao2025robust} &
I &
85.1K &
54 &
\yes & \yes & & &
\yes & \yes & \\

\rowcolor{seedHL}
\textbf{SEED (Ours)} &
\textbf{I} &
\textbf{91.5K} &
\textbf{200} &
\textbf{\yes} & \textbf{\yes} & \textbf{\yes} & \textbf{\yes} &
\textbf{\yes} & \textbf{\yes} & \textbf{\yes} \\

\bottomrule
\end{tabular}

\vspace{0.15em}
{\scriptsize
\textit{Supervision:} Ord = edit order; Attr = attribute label; Txt = text instruction; Mask = spatial mask.

\quad \textit{Tasks:} Det = detection; Prov = provenance tracing; Loc = localization.

\quad \textit{Seq:} number of sequence types; ``-'' indicates not applicable.

\quad $^{\dagger}$GenImage reports $>$1M real-fake pairs.}
\endgroup
\end{table}

\section{Introduction}

Recent progress in diffusion models has enabled highly realistic and controllable face synthesis and editing at scale~\cite{rombach2022latentdiffusion,podell2023sdxl,esser2024scaling}.
While these tools benefit content creation, they also blur the boundary between authentic and manipulated imagery, amplifying the need for reliable visual provenance analysis~\cite{rossler2019faceforensicspp,jiang2020deeperforensics}.
Importantly, practical editing workflows are rarely single-shot.
Users can perform \emph{sequential} attribute edits (e.g., hair, eyes, eyebrows, accessories), producing a final image that reflects an ordered chain of operations.
Such multi-step pipelines are particularly challenging for forensics because evidence is \emph{non-stationary}: later edits may overwrite earlier traces, distributed, and entangled across attributes.

Despite strong progress in face forensics benchmarks and detectors, existing datasets predominantly emphasize isolated manipulations or global authenticity labels.
Classic benchmarks such as FaceForensics++~\cite{rossler2019faceforensicspp} and DeeperForensics-1.0~\cite{jiang2020deeperforensics} focus on detection under controlled or real-world distortions.
Large-scale datasets such as ForgeryNet~\cite{he2021forgerynet} and OpenForensics~\cite{le2021openforensics} provide rich supervision for detection, but do not model \emph{editing histories}.
More recently, diffusion-era resources and detectors (e.g., diffusion-generated detection via reconstruction error~\cite{wang2023dire} and diffusion-focused datasets~\cite{chen2024diffusionface}) highlight the new threat landscape, yet still largely assume \emph{single-edit} settings.
The only widely-used sequential benchmark, Seq-DeepFake~\cite{shao2022sequential}, is built upon GAN-based pipelines, which limits its coverage of diffusion-based sequential edits.
This gap makes it difficult to systematically study how provenance cues evolve across multiple diffusion edits, and how to recover attribute identity and temporal order from subtle evidence.

To address this need, we introduce \underline{SE}quential \underline{E}diting in \underline{D}iffusion (\textbf{SEED}), the first large-scale benchmark for diffusion-based sequential provenance tracing in facial imagery.
SEED contains 91,526 face images generated by applying one to four attribute edits sequentially, using modern diffusion editing pipelines such as LEDITS~\cite{tsaban2023ledits}, SDXL~\cite{podell2023sdxl}, and SD3~\cite{esser2024scaling} style rectified-flow models~\cite{esser2024scaling}.
Crucially, each sample is paired with \emph{step-wise provenance metadata}, including the edited attribute sequence, textual instructions, and manipulation masks.
As summarized in Figure~\ref{fig:seed_overview} and Table~\ref{tab:forgery_datasets_tick}, this design enables three complementary tasks:
(i) \emph{Authenticity Analysis} (real vs. sequentially edited),
(ii) \emph{Editing Trace Analysis} (predicting edited attributes and their order),
and (iii) \emph{Spatial Evidence Analysis} (localizing manipulated regions).

We further benchmark representative analysis strategies on SEED and observe that spatial-only approaches are often insufficient under subtle diffusion artifacts, especially as the edit chain becomes longer.
Motivated by findings that mid-to-high frequency behaviors remain informative for synthetic imagery~\cite{corvi2023intriguing,tan2024upsampling}, we establish \textbf{FAITH} as a frequency-aware Transformer \emph{diagnostic baseline} for sequential provenance tracing.
FAITH is not proposed as a state-of-the-art solution, but is designed to isolate and study the contribution of frequency-domain signals when recovering attribute identity and temporal order under multi-step diffusion edits.
Concretely, FAITH aggregates spatial features with frequency-domain cues to support controlled analyses of which signals persist, degrade, or become confounded as edits accumulate. In summary, our contributions are threefold:
\begin{itemize}[leftmargin=*]
    \vspace{-3mm}
    \item We present \textbf{SEED}, the first large-scale benchmark for diffusion-based sequential provenance tracing in facial imagery, comprising 91,526 images with step-wise annotations (edit order, attribute labels, prompts, and masks) for explainable analysis.
    \item We provide systematic benchmarking on SEED, highlighting the difficulty of recovering complete edit histories under multi-step diffusion edits and motivating evidence aggregation beyond purely spatial cues.
    \item We establish \textbf{FAITH} as a frequency-aware diagnostic baseline that integrates spatial and frequency cues for sequential provenance tracing, enabling controlled studies of robustness and signal contribution.
\end{itemize}

\section{Related Work}

\noindent\textbf{Face Forensics Benchmarks and Supervision.}
Early and large-scale benchmarks have driven progress in deepfake detection.
FaceForensics++~\cite{rossler2019faceforensicspp} standardized evaluation for manipulated facial videos, while DeeperForensics-1.0~\cite{jiang2020deeperforensics} emphasized scale and real-world perturbations.
Celeb-DF~\cite{li2020celeb} improved visual quality for face-swapping evaluation.
Beyond global labels, richer supervision has been introduced via mega-scale resources such as ForgeryNet~\cite{he2021forgerynet} and OpenForensics~\cite{le2021openforensics}, and DF-Platter~\cite{narayan2023dfplatter} further highlights real-world variability, including multi-subject settings.
These datasets enable comprehensive detection studies and motivate more generalizable and interpretable detectors~\cite{lin2024cdfa,tian2024ram,zhang2024naco,zhang2024commonsense}, but they largely treat each sample as a single manipulation outcome and do not explicitly represent \emph{multi-step editing histories}.
Diffusion-era resources have recently emerged as diffusion models become dominant for synthesis and editing.
DiffusionDB~\cite{wang2022diffusiondb} provides large-scale prompt-image pairs for text-to-image generation analysis, and DiffusionFace~\cite{chen2024diffusionface} targets diffusion-based face forgery analysis, together with diffusion-oriented representation learning and prompt-based forensics efforts~\cite{baraldi2024code,li2024noise_prompt}.
However, these datasets primarily support detection or categorization under \emph{single-generation/single-edit} assumptions.
In contrast, SEED focuses on diffusion-based \emph{sequential} edits and records step-wise provenance (order, prompts, masks), enabling provenance tracing beyond binary authenticity labels.

\noindent\textbf{Diffusion and Sequential Forensics.}
For diffusion-generated imagery, recent work explores signals that remain discriminative despite improved realism.
DIRE~\cite{wang2023dire} proposes diffusion reconstruction error for diffusion-generated image detection, while analyses of synthetic image properties show systematic mid-to-high frequency irregularities across generators~\cite{corvi2023intriguing,qian2020f3net,liu2022realimages}.
Generalizable deepfake detection has also been advanced by explicitly modeling generator-induced artifacts, including up-sampling traces~\cite{tan2024upsampling} and localized artifact attention~\cite{nguyen2024laanet,zhuang2022uiavit,gu2022hcil}, and related image manipulation detection under limited supervision has also been explored~\cite{zhang2024inr_imd}.
Beyond detection, weakly-supervised localization for diffusion-generated images has been studied to provide spatial evidence maps rather than only binary predictions~\cite{tantaru2024weakly,sun2022infotheory}.
Sequential provenance tracing is inherently more challenging than single-step detection: as edits accumulate, the model must determine not only which attributes were modified, but also the order in which those modifications occurred.
Seq-DeepFake~\cite{shao2022sequential} formulates sequential manipulation as an image-to-sequence problem and provides step-wise order supervision, and subsequent work improves robustness under post-processing perturbations~\cite{shao2025robust}.
Nevertheless, existing sequential benchmarks and methods are largely built on GAN-era editing pipelines, and their coverage of diffusion-based sequential edits remains limited.
SEED complements this line by providing a diffusion-era sequential benchmark with fine-grained, step-wise provenance annotations for detection.
\begin{figure}[!t]
  \centering
  \includegraphics[width=\linewidth]{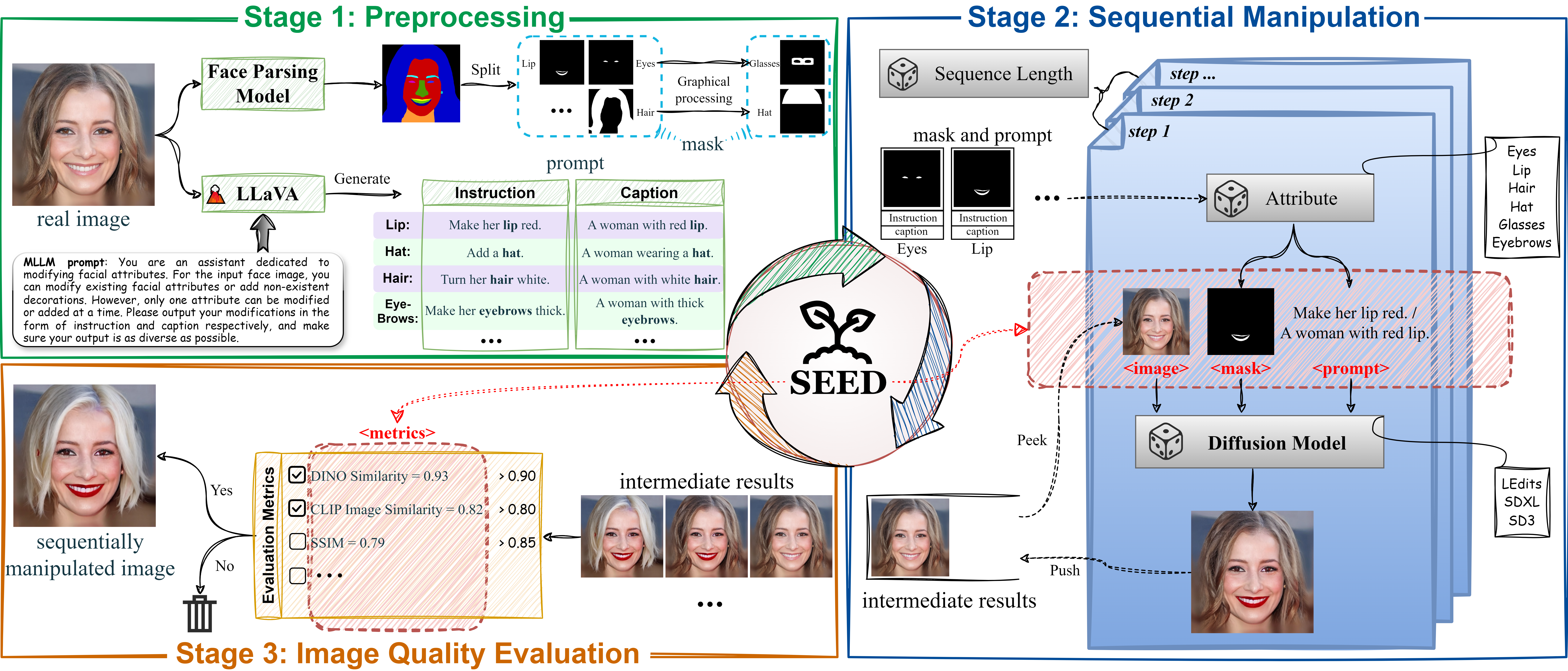}
  \caption{\textbf{Overview of the SEED dataset construction pipeline.}
  We generate masks and prompts, apply step-wise diffusion editing with per-step model sampling, and finally compute quality metrics while recording step-wise provenance annotations.}
  \label{fig:seed_pipeline}
\end{figure}
\vspace{-3mm}
\section{SEED: A Benchmark for Sequential Visual Provenance Tracing}
\label{sec:seed}

We introduce \textbf{SEED}, the first large-scale benchmark for \emph{sequential} visual provenance tracing in facial imagery.
SEED contains 91{,}526 images built from high-quality real face datasets (FFHQ~\cite{karras2019stylegan} and CelebAMask-HQ~\cite{lee2020maskgan}) and edited by diffusion-based pipelines.
Unlike prior benchmarks that focus on isolated manipulations, SEED targets \emph{sequential} editing: each sample is produced by applying one to four attribute edits in order, where later operations may partially overwrite earlier traces.
For each step, we record provenance annotations including the edited attribute, the textual instruction, the spatial mask, and the editing model, yielding trajectory-level supervision (the edit sequence) together with step-level evidence for analysis.

\begin{SCfigure}[0.35][t]
  \centering
  \includegraphics[width=0.62\linewidth]{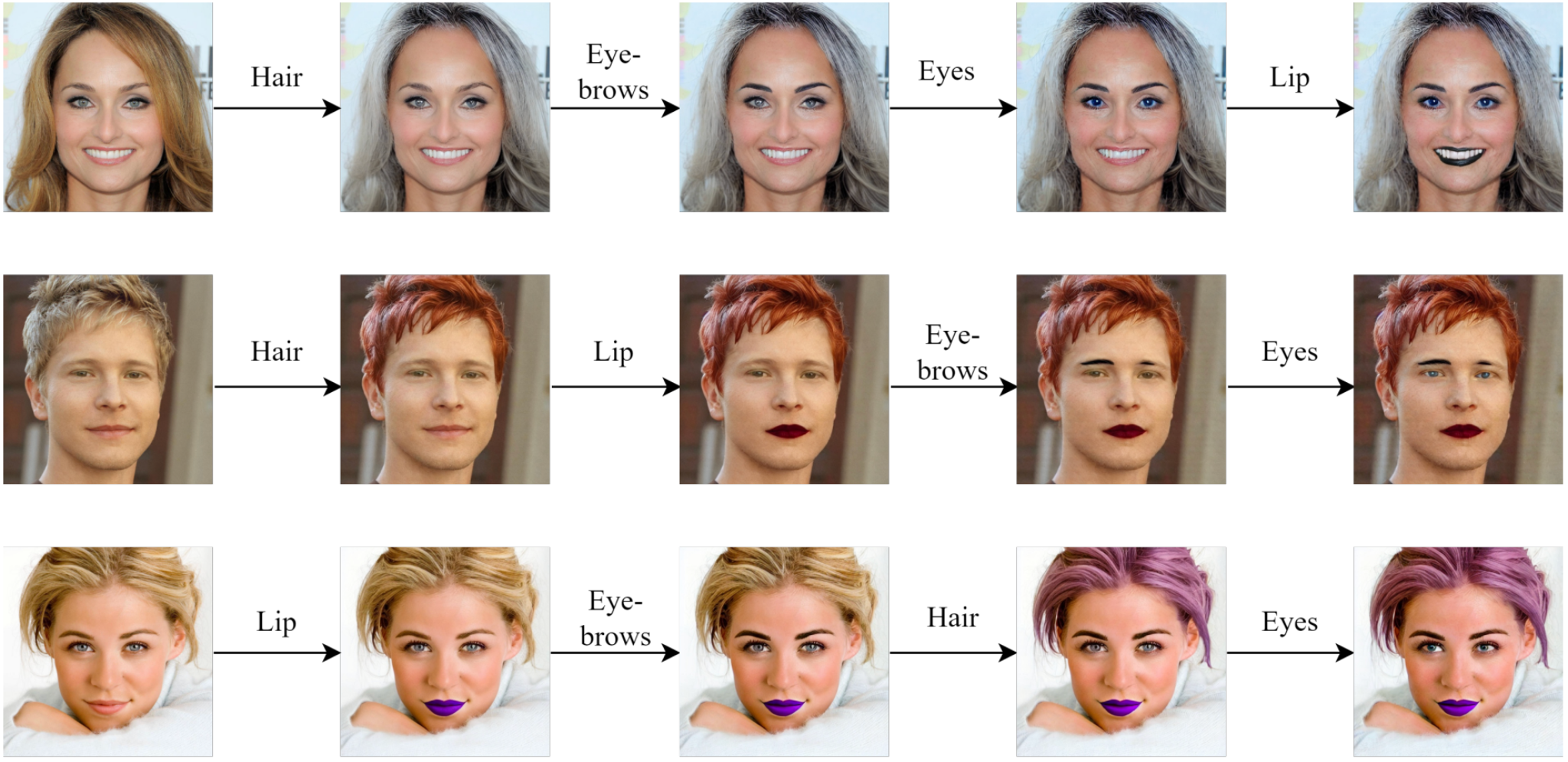}
  \caption{\textbf{Sequential edit examples and edit order.}
  Each row shows a multi-step editing trajectory in SEED.}
  \label{fig:seed_edit_order}
  \vspace{-2mm}
\end{SCfigure}

\vspace{-2mm}
\subsection{Dataset Construction}
\vspace{-2mm}
\label{subsec:seed_construction}

SEED is constructed via three stages: preprocessing, sequential manipulation, and quality evaluation (Figure~\ref{fig:seed_pipeline}).
\vspace{-2mm}
\paragraph{Preprocessing: spatial and textual conditions.}
We build attribute-specific spatial and textual conditions for localized, instruction-driven edits.
For intrinsic attributes (eyes, lips, hair, eyebrows), we obtain region masks using CelebAMask-HQ parsing~\cite{lee2020maskgan}.
For accessories (glasses, hats), we use scalable geometric priors (e.g., dilated eye/upper-face regions).

To introduce linguistic variation without changing edit intent, we generate multiple candidate prompts per attribute using LLaVA~\cite{liu2023visual} and randomly sample instruction-style or description-style templates per step.
Table~\ref{tab:seed_prompt_templates} summarizes representative prompt templates used in SEED data generation.

\begin{table}[t!]
\centering
\scriptsize
\setlength{\tabcolsep}{4.5pt}
\renewcommand{\arraystretch}{1.15}
\caption{\textbf{Prompt templates used in SEED data generation.}
Curly braces denote placeholders sampled from predefined vocabularies to diversify linguistic realizations while keeping edit intent fixed.}
\label{tab:seed_prompt_templates}
\begin{adjustbox}{max width=\linewidth,center}
\begin{tabular}{l p{0.36\linewidth} p{0.36\linewidth} p{0.20\linewidth}}
\toprule
\textbf{Attribute} & \textbf{Instruction template} & \textbf{Caption template} & \textbf{Placeholders} \\
\midrule
Eyes
& Make the eyes \{color\}.
& A person with \{color\} eyes.
& red, blue, black, \dots \\

Lip
& Change the lipstick color to \{color\}.
& A person with \{color\} lipstick.
& red, blue, black, \dots \\

\multirow{2}{*}{Hair}
& Turn the hair \{color\}.
& A person with \{color\} hair.
& red, blue, black, \dots \\
& Make the hair \{style\}.
& A person with \{style\} hair.
& curly, straight, \dots \\

Eyebrows
& Make the eyebrows \{style\}.
& A person with \{style\} eyebrows.
& thick, \dots \\

Glasses
& Add a pair of \{glasses\}.
& A person wearing \{glasses\}.
& glasses, sunglasses \\

Hat
& Add a \{hat\}.
& A person wearing a \{hat\}.
& hat, \dots \\
\bottomrule
\end{tabular}
\end{adjustbox}
\end{table}

\paragraph{Sequential manipulation.}
We sample the sequence length $L \in \{1,2,3,4\}$, choose an attribute at each step, and apply a diffusion editor.
To diversify editor-specific behaviors, we sample the per-step model from a pool including LEdits~\cite{tsaban2023ledits}, SDXL~\cite{podell2023sdxl}, and SD3~\cite{esser2024scaling} fine-tuned with UltraEdit~\cite{zhao2024ultraedit}.
The output of each step becomes the input to the next, forming a cumulative editing chain with explicit provenance logs.

\paragraph{Quality evaluation.}
We filter degenerate outputs and retain trajectories with reasonable semantic and structural consistency.
We compute perceptual and semantic similarity measurements to avoid obvious collapses, and store the raw scores as metadata for subsequent analysis.
Detailed definitions, thresholds, and the full metadata schema are provided in the supplementary.

\subsection{Dataset Distributions}
\label{subsec:seed_stats}

\begin{figure}[t]
  \centering
  \includegraphics[width=0.9\linewidth]{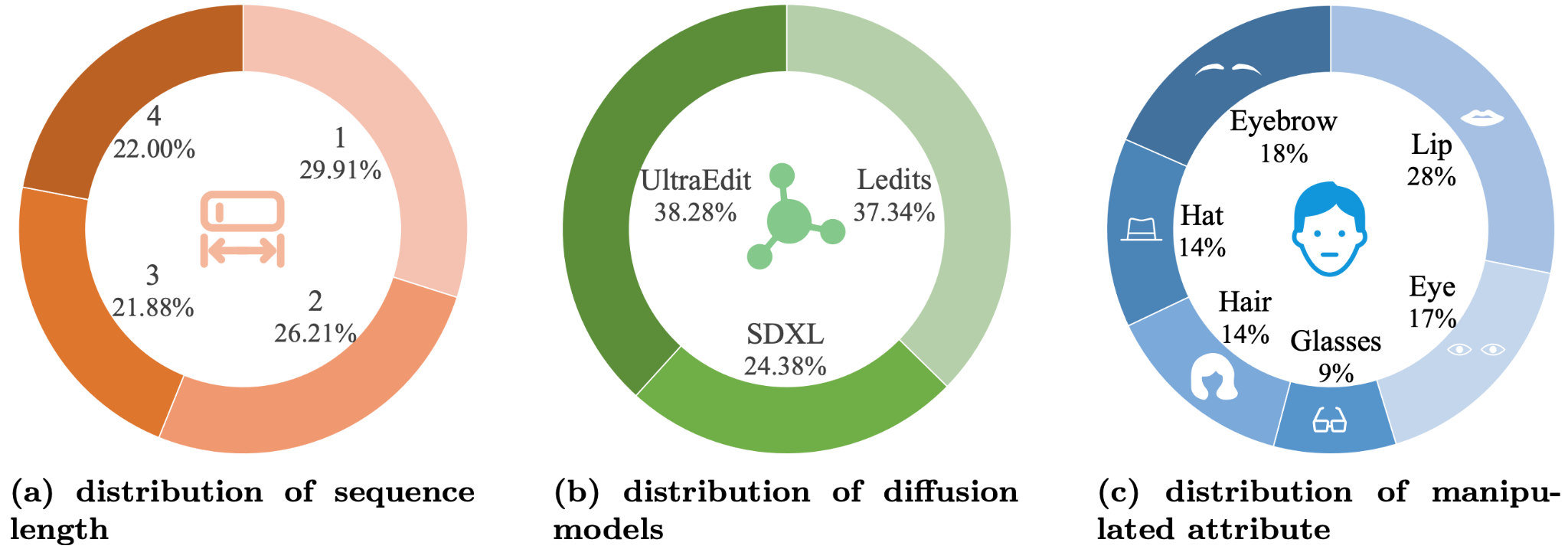}
  \caption{\textbf{SEED dataset distributions.}
  (a) Distribution of sequence length.
  (b) Distribution of diffusion editing models.
  (c) Distribution of manipulated attribute categories.}
  \label{fig:seed_distribution}
  \vspace{-.1in}
\end{figure}

Figure~\ref{fig:seed_distribution} summarizes SEED statistics from three perspectives.
(1) The sequence-length distribution is relatively balanced, with 29.91\%, 26.21\%, 21.88\%, and 22.00\% of samples having $L=1,2,3,4$ edits, respectively (Fig.~\ref{fig:seed_distribution}a).
(2) SEED mixes multiple diffusion editors to discourage over-specialization to a single generator (Fig.~\ref{fig:seed_distribution}b), including UltraEdit (38.28\%), LEdits (37.34\%), and SDXL (24.38\%).
(3) SEED covers both intrinsic facial attributes and accessories (Fig.~\ref{fig:seed_distribution}c), including Lip (28\%), Eyebrow (18\%), Eye (17\%), Hat (14\%), Hair (14\%), and Glasses (9\%).

\subsection{Balanced Partition and Split Protocol}
\label{subsec:seed_split}

\paragraph{Identity-disjoint split.}
SEED adopts an identity-disjoint split for training, validation, and testing, where identities do not overlap across splits.
This reduces identity leakage and encourages models to rely on provenance cues rather than subject memorization.

\paragraph{Sequence-length-balanced partition.}
Sequential provenance tracing becomes progressively harder as edit chains grow, therefore length imbalance can bias both training and evaluation.
To control for this factor, we construct a length-balanced partition for benchmarking.
Specifically, we sample 20{,}000 images for each edited length $L \in \{1,2,3,4\}$ and additionally include 20{,}000 unedited real images ($L=0$) from FFHQ and CelebAMask-HQ, yielding 100{,}000 images in total with each length accounting for 20\%.
We then split this partition into train/val/test sets with an 8:1:1 ratio while preserving the length distribution.

\begin{figure*}[htbp]
  \centering
  \includegraphics[width=\linewidth]{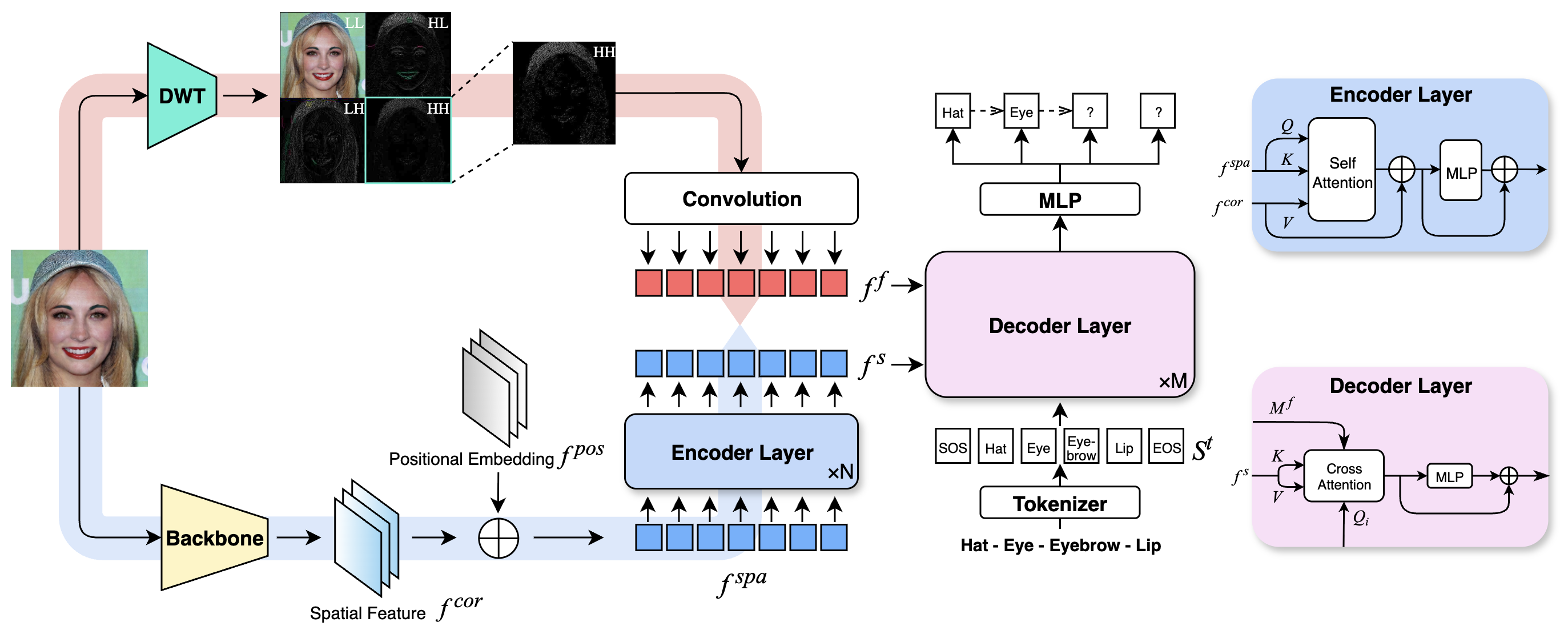}
  \caption{Overview of the proposed FAITH architecture. The model explicitly integrates high-frequency domain information extracted by Discrete Wavelet Transform (DWT) and spatial-domain features via a Transformer-based encoder-decoder structure, significantly enhancing sequential facial editing detection accuracy.}
  \label{Sequential deepfake detection model}
\vspace{-.1in}
\end{figure*}

\section{FAITH: Frequency-Aware Identification Transformer for Sequential Editing Detection}

Existing sequential facial editing detectors, such as SeqFakeFormer~\cite{shao2022sequential}, mainly rely on spatial-domain representations.
This design can be insufficient for subtle and realistic diffusion-based edits~\cite{ricker2022towards}, where spatial cues are weak and accumulated artifacts are distributed.
Moreover, attention maps derived solely from label embeddings may fail to consistently highlight manipulated regions, especially under multi-step edits.
To address these limitations, we propose \textbf{FAITH}, a frequency-aware encoder-decoder Transformer baseline for sequential editing detection on SEED.
As illustrated in Figure~\ref{Sequential deepfake detection model}, FAITH couples spatial features with high-frequency guidance extracted by DWT, and injects such guidance into the decoder cross-attention to improve sensitivity to subtle editing traces.

\noindent\textbf{Problem Setup and Notation.}
Given an input face image $\mathbf{x}\in\mathbb{R}^{H\times W\times 3}$, the goal is to predict an \emph{ordered} attribute sequence that describes the sequential edits (up to four steps in SEED).
Let the ground-truth edit sequence be $\mathbf{y}=(y_1,\ldots,y_T)$ with length $T\in\{0,1,2,3,4\}$, where each $y_i$ is an attribute token from the vocabulary $\mathcal{V}$ (e.g., Hat, Eye, Eyebrow, Lip).
We prepend a start token $\langle\mathrm{SOS}\rangle$ and append an end token $\langle\mathrm{EOS}\rangle$ to indicate termination, yielding $(\langle\mathrm{SOS}\rangle, y_1,\ldots,y_T,\langle\mathrm{EOS}\rangle)$.
When $T=0$ (no edit), the target sequence naturally reduces to $\langle\mathrm{SOS}\rangle \rightarrow \langle\mathrm{EOS}\rangle$, which avoids introducing a separate ``real'' label outside the sequence formulation.

\subsection{Spatial Feature Extraction and Encoding}
\noindent\textbf{Spatial tokenization.}
FAITH first extracts coarse spatial features $f^{\mathrm{cor}}$ via a CNN backbone, then adds positional embeddings and flattens them into a token sequence $f^{\mathrm{spa}}$:
\begin{equation}
\begin{aligned}
f^{\mathrm{cor}} &= \mathrm{CNN}(\mathbf{x}),\quad
f^{\mathrm{spa}} &= \mathrm{Flatten}\!\left(f^{\mathrm{cor}} + f^{\mathrm{pos}}\right),
\end{aligned}
\label{eq:faith_spatial_tokenize}
\end{equation}
where $f^{\mathrm{pos}}$ denotes the learnable positional embedding broadcastable to $f^{\mathrm{cor}}$.
After flattening, $f^{\mathrm{spa}}$ is a sequence of $L$ spatial tokens.

\noindent\textbf{Transformer encoder.}
We refine $f^{\mathrm{spa}}$ with an $N$-layer Transformer encoder to obtain refined spatial features $f^{\mathrm{s}}$.
For clarity, we write one encoder layer using the standard attention-then-MLP residual form:
\begin{equation}
\begin{aligned}
Q &= f^{\mathrm{spa}} W_Q,\quad K = f^{\mathrm{spa}} W_K,\quad V = f^{\mathrm{spa}} W_V,\\
f^{\mathrm{mid}} &= f^{\mathrm{spa}} + \mathrm{SoftMax}\!\left(\frac{QK^{\top}}{\sqrt{d}}\right)V,\\
f^{\mathrm{s}} &= f^{\mathrm{mid}} + \mathrm{MLP}(f^{\mathrm{mid}}),
\end{aligned}
\label{eq:faith_encoder_layer}
\end{equation}
where $W_Q,W_K,W_V\in\mathbb{R}^{d\times d}$ are learnable projections (applied per head in the multi-head implementation), and $d$ is the hidden dimension.
Stacking $N$ layers yields the final spatial representation $f^{\mathrm{s}}$, which serves as the encoder memory for decoding.

\subsection{Frequency-domain Feature Extraction and Guidance Construction}
\noindent\textbf{DWT-based high-frequency features.}
To explicitly highlight subtle manipulation traces, FAITH leverages the discrete wavelet transform (DWT) and focuses on the HH sub-band, which captures high-frequency details along both horizontal and vertical directions.
We encode HH with a lightweight convolutional head to obtain frequency features $f^{f}$:
\begin{equation}
\{\mathrm{LL}, \mathrm{LH}, \mathrm{HL}, \mathrm{HH}\} = \mathrm{DWT}(\mathbf{x}),
\quad
f^{f} = \mathrm{Conv}(\mathrm{HH}).
\label{eq:faith_freq_feature}
\end{equation}
where $\mathrm{HH}$ denotes the extracted high-frequency component after wavelet decomposition.

\noindent\textbf{Frequency guidance map.}
To match the decoder cross-attention operation in Figure~\ref{Sequential deepfake detection model}, we convert $f^{f}$ into an additive guidance term $M^{f}$ that modulates attention scores over spatial tokens.
Concretely, we align $f^{f}$ to the spatial-token resolution and project it to an attention bias:
\begin{equation}
\begin{aligned}
M^{f} &= \mathrm{Flatten}\!\left(\mathrm{Proj}\!\left(\mathrm{Align}(f^{f})\right)\right),
\end{aligned}
\label{eq:faith_freq_guidance}
\end{equation}
where $\mathrm{Align}(\cdot)$ denotes resolution alignment (e.g., interpolation to the encoder token grid), and $\mathrm{Proj}(\cdot)$ denotes a learnable projection (e.g., $1\times1$ convolution) to produce a token-wise bias compatible with attention logits.
This design preserves the figure-level semantics: $f^{f}$ is computed from HH, and $M^{f}$ is the guidance injected into cross-attention.

\subsection{Cross-attention based Sequential Edit Prediction}
\noindent\textbf{Tokenization and decoder input.}
We tokenize each edited attribute and insert special tokens to form the target sequence.
Let $S^{\mathrm{t}}=(s_0,s_1,\ldots,s_T,s_{T+1})$ denote the token sequence where $s_0=\langle\mathrm{SOS}\rangle$ and $s_{T+1}=\langle\mathrm{EOS}\rangle$.
The decoder maintains an auto-regressive hidden state and predicts tokens step by step.

\noindent\textbf{Frequency-aware cross-attention.}
At decoding step $i$, let the decoder produce a query representation (after its self-attention) denoted as $\mathbf{u}_i$.
We compute cross-attention between $\mathbf{u}_i$ (query) and the refined spatial features $f^{\mathrm{s}}$ (keys and values), while injecting $M^{f}$ as an additive bias to guide attention toward high-frequency suspicious regions:
\begin{equation}
\begin{aligned}
Q_i &= \mathbf{u}_i W_Q^{\mathrm{d}},\quad
K = f^{\mathrm{s}} W_K^{\mathrm{d}},\quad
V = f^{\mathrm{s}} W_V^{\mathrm{d}},\\
f^{\mathrm{mid}'}_i &= \mathrm{SoftMax}\!\left(\frac{Q_i K^{\top}}{\sqrt{d}} + M^{f}\right) V,\\
f^{\mathrm{dec}}_i &= f^{\mathrm{mid}'}_i + \mathrm{MLP}\!\left(f^{\mathrm{mid}'}_i\right),
\end{aligned}
\label{eq:faith_decoder_cross_attn}
\end{equation}
where $W_Q^{\mathrm{d}},W_K^{\mathrm{d}},W_V^{\mathrm{d}}\in\mathbb{R}^{d\times d}$ are learnable projections for the decoder cross-attention.
The guidance term $M^{f}$ is shared across decoding steps and biases the attention distribution over spatial tokens.

\noindent\textbf{Attribute prediction and termination.}
The predicted attribute distribution at step $i$ is computed by a linear classifier followed by SoftMax:
\begin{equation}
\begin{aligned}
p_{\theta}(s_i \mid s_{<i}, \mathbf{x})
&= \mathrm{SoftMax}\!\left(f^{\mathrm{dec}}_i W_O\right),
\end{aligned}
\label{eq:faith_token_prediction}
\end{equation}
where $W_O\in\mathbb{R}^{d\times|\mathcal{V}\cup\{\langle\mathrm{EOS}\rangle\}|}$.
Inference proceeds auto-regressively until $\langle\mathrm{EOS}\rangle$ is generated, or a maximum length is reached (four edit tokens plus $\langle\mathrm{EOS}\rangle$), consistent with the SEED protocol.

\subsection{Training Objective}
We train FAITH with teacher forcing using the standard cross-entropy loss over the full token sequence (including $\langle\mathrm{EOS}\rangle$):
\begin{equation}
\begin{aligned}
\mathcal{L}
&= - \sum_{i=1}^{T+1} \log p_{\theta}(s_i^{*}\mid s_{<i}^{*}, \mathbf{x}),
\end{aligned}
\label{eq:faith_loss}
\end{equation}
where $s_i^{*}$ denotes the ground-truth token at position $i$.
This objective directly supervises both the ordered attribute predictions and the termination behavior, enabling FAITH to model sequential edits under a unified formulation while leveraging $M^{f}$ for frequency-aware detection cues during decoding.
During training, the decoder conditions on the ground-truth prefix $s_{<i}^{*}$, which stabilizes optimization and encourages each next-token distribution to be locally consistent with the true edit history.
At inference time, we decode autoregressively from $\langle\mathrm{BOS}\rangle$ until $\langle\mathrm{EOS}\rangle$ (or the maximum length), allowing FAITH to recover edit order and termination jointly while leveraging the frequency-aware branch $M^{f}$ to provide complementary cues when spatial traces are subtle or partially overwritten.
\section{Experiment}
\subsection{Implementation Details}

For FAITH, we use a ResNet-50 backbone pre-trained on ImageNet for spatial feature extraction.
The encoder-decoder Transformer contains 2 encoder layers and 2 decoder layers, each with 4 attention heads.
We train for 170 epochs using a 20-epoch warm-up, followed by step-wise learning rate decay every 50 epochs.
The initial learning rates are $1\times10^{-3}$ for the Transformer and $1\times10^{-4}$ for the ResNet backbone, and the batch size is 40 with sequence padding up to length 4.
All experiments are implemented in PyTorch and run on three NVIDIA RTX 3090 GPUs, and we adopt SAM to improve generalization.
For comparison, we adapt representative face manipulation detectors from binary classification to a 200-way multi-class setting, assessing how well non-sequential designs transfer to sequential provenance tracing.
We further include SeqFakeFormer as a sequential baseline that relies only on spatial cues, isolating the benefit of explicitly incorporating frequency-domain information in FAITH.

\begin{table*}[t]
\centering
\small
\renewcommand{\arraystretch}{1.05}
\setlength{\tabcolsep}{4.5pt}
\caption{
Performance on SEED across sequence length $L$ (number of sequential edits).
We compare adapted manipulation detectors, a sequential baseline, and FAITH with frequency transforms (DCT, FFT, DWT).
We report Fixed-Acc, Adaptive-Acc, and Full-Acc, where Full-Acc requires the entire predicted edit sequence to exactly match the ground truth.
Best and second-best results in each row are marked by \best{bold} and \second{underline}, respectively.
FAITH columns are highlighted in light blue.
}
\label{tab:methods_comparison}

\begin{adjustbox}{max width=0.98\textwidth,center}
\begin{tabular}{cc|ccc|c|>{\columncolor{faithblue}}c>{\columncolor{faithblue}}c>{\columncolor{faithblue}}c}
\toprule
\multicolumn{2}{c|}{\textbf{Settings}} &
Shuai \textit{et al.}~\cite{shuai2023locateverify} &
FreqNet~\cite{tan2024freqnet} &
Ba \textit{et al.}~\cite{ba2024exposingdeception} &
SeqFakeFormer~\cite{shao2022sequential} &
\textbf{FAITH(DCT)} &
\textbf{FAITH(FFT)} &
\textbf{FAITH(DWT)} \\
\midrule

\multirow{2}{*}{$L=0$}  & \textit{Fixed}    & 99.95 & \best{100} & \second{99.99} & \best{100} & \best{100} & \best{100} & \best{100} \\
                       \multirow{2.5}{*}{(no edits)}& \textit{Adaptive} & 99.80 & \best{100} & \second{99.95} & \best{100} & \best{100} & \best{100} & \best{100} \\
                       & \textit{Full}     & 99.80 & \best{100} & \second{99.95} & \best{100} & \best{100} & \best{100} & \best{100} \\
\midrule

\multirow{3}{*}{$L=1$}  & \textit{Fixed}    & 97.32 & 93.77 & 93.25 & 98.00 & 98.10 & \best{98.29} & \second{98.24} \\
                       & \textit{Adaptive} & 90.15 & 80.37 & 87.98 & 92.80 & 93.10 & \best{93.60} & \second{93.40} \\
                       & \textit{Full}     & 89.50 & 79.01 & 86.61 & 92.11 & 92.48 & \best{93.21} & \second{93.01} \\
\midrule

\multirow{3}{*}{$L=2$}  & \textit{Fixed}    & 78.12 & 74.01 & 68.01 & 89.50 & \best{89.76} & \second{89.70} & 89.69 \\
                       & \textit{Adaptive} & 56.88 & 49.82 & 48.62 & 79.44 & \best{79.82} & \second{79.62} & \second{79.62} \\
                       & \textit{Full}     & 52.25 & 46.09 & 52.31 & 76.75 & \best{77.10} & \second{77.00} & \best{77.10} \\
\midrule

\multirow{3}{*}{$L=3$}  & \textit{Fixed}    & 53.66 & 51.34 & 52.60 & 70.44 & 71.12 & \second{71.61} & \best{72.50} \\
                       & \textit{Adaptive} & 35.53 & 32.54 & 40.62 & 61.56 & 62.21 & \second{62.67} & \best{63.84} \\
                       & \textit{Full}     & 22.45 & 25.57 & 26.58 & 48.55 & \second{49.65} & 49.25 & \best{50.35} \\
\midrule

\multirow{3}{*}{$L=4$}  & \textit{Fixed}    & 28.45 & 26.93 & 34.20 & \best{50.14} & \second{49.51} & 49.14 & 48.93 \\
                       & \textit{Adaptive} & 28.45 & 26.93 & 34.20 & \best{50.14} & \second{49.51} & 49.14 & 48.93 \\
                       & \textit{Full}     & 5.70  & 10.92 & 7.20  & \best{24.55} & 22.95 & 23.05 & \second{24.50} \\
\midrule

\multirow{3}{*}{Avg.}   & \textit{Fixed}    & 71.50 & 70.08 & 68.78 & 81.62 & 81.70 & \second{81.75} & \best{81.87} \\
                       & \textit{Adaptive} & 54.07 & 52.59 & 54.80 & 68.53 & 68.56 & \second{68.58} & \best{68.84} \\
                       & \textit{Full}     & 48.72 & 48.27 & 50.80 & 66.97 & 67.02 & \second{67.03} & \best{67.26} \\
\bottomrule
\end{tabular}
\end{adjustbox}
\end{table*}

\subsection{Results and Analysis}
We study whether frequency-domain cues help recover sequential facial edit histories.
To this end, we instantiate FAITH with three standard transforms (DCT, FFT, DWT) and compare them with adapted manipulation detectors and a sequential baseline.
Table~\ref{tab:methods_comparison} reports results across sequence length $L$ and three complementary metrics (Fixed-Acc, Adaptive-Acc, and Full-Acc).

Two trends are consistent throughout the table.
First, detectors designed for single-step manipulation show limited transfer to sequential provenance tracing.
While they saturate at $L{=}0$, their performance drops substantially once the task requires predicting an \emph{ordered} edit sequence.
This indicates that sequential provenance is not merely ``detecting edits'', but separating multiple attribute traces that can be weak or partially overwritten by later operations.

Second, accuracy decreases as $L$ increases for all approaches, and the drop is most visible under Full-Acc.
Because Full-Acc requires an exact match of the entire sequence, any step-wise mistake (wrong attribute, wrong order, or early stopping) collapses the score.
In contrast, Fixed-Acc and Adaptive-Acc allow partial credit at the token level, so they degrade more gently, suggesting that many predictions remain partially correct.

Within sequential methods, adding frequency-domain features improves FAITH across metrics in most settings.
The gains are most meaningful on Full-Acc, implying that frequency cues help reduce hard sequence-level failures such as ordering drift or missing steps.
Among the three transforms, DWT achieves the highest average performance.
Compared to DCT/FFT, DWT provides localized high-frequency decomposition, which better preserves edge and texture changes introduced by attribute edits, especially when multiple edits accumulate.
This interpretation is supported by the qualitative robustness examples in Fig.~6, where our method maintains the full edit sequence under Compress-50\% and Noise intensity-15\%, while baselines more often terminate early or deviate from the correct order.
At the same time, $L{=}4$ remains difficult for all methods and the margin between strong sequential models narrows, highlighting that long edit chains still pose substantial challenges.

\begin{table*}[t]
\centering
\small
\renewcommand{\arraystretch}{1.05}
\setlength{\tabcolsep}{3.8pt}
\caption{Robustness on the SEED test set under two common post-processing conditions: (a) JPEG compression (25\%, 50\%, 75\%) and (b) Gaussian noise (10\%, 15\%, 20\%).
We report Fixed-Acc, Adaptive-Acc, and Full-Acc.
Best and second-best results in each column are marked by \best{bold} and \second{underline}, respectively.
FAITH variants are highlighted in light blue and separated from prior baselines by a horizontal rule.}

\scriptsize
\label{tab:robustness}
\vspace{-5mm}
\begin{subtable}[t]{\linewidth}
\centering
\caption{}
\vspace{-.1in}
\label{tab:jpeg}
\begin{adjustbox}{max width=0.98\linewidth,center}
\begin{tabular}{l *{3}{ccc}}
\toprule
\multirow{2}{*}{\textbf{Methods}} &
\multicolumn{3}{c}{Compress -- 25\%} &
\multicolumn{3}{c}{Compress -- 50\%} &
\multicolumn{3}{c}{Compress -- 75\%} \\
\cmidrule(lr){2-4} \cmidrule(lr){5-7} \cmidrule(lr){8-10}
& \textit{Fixed} & \textit{Adaptive} & \textit{Full} &
\textit{Fixed} & \textit{Adaptive} & \textit{Full} &
\textit{Fixed} & \textit{Adaptive} & \textit{Full} \\
\midrule
Shuai \textit{et al.}~\cite{shuai2023locateverify} & 63.98 & 50.21 & 47.81 & 61.06 & 48.68 & 46.29 & 56.87 & 36.22 & 32.69 \\
FreqNet~\cite{tan2024freqnet}                        & 63.20 & 50.03 & 47.09 & 60.73 & 48.70 & 46.04 & 56.41 & 36.43 & 32.44 \\
Ba \textit{et al.}~\cite{ba2024exposingdeception}
                                              & 63.15 & 50.36 & 47.47 & 59.93 & 48.98 & 45.90 & 56.76 & 36.41 & 33.01 \\
SeqFakeFormer~\cite{shao2022sequential}
                                              & 78.12 & 63.67 & 60.60 & 75.83 & 62.01 & 59.24 & 66.63 & 44.41 & 39.33 \\
\midrule
\rowcolor{faithblue}
FAITH (DCT) & 78.68 & 63.93 & 61.41 & \second{76.89} & 62.09 & \second{59.78} & 66.80 & 44.36 & 39.18 \\
\rowcolor{faithblue}
FAITH (FFT) & \second{78.83} & \second{63.98} & \second{61.43} & 76.64 & \second{62.10} & 59.37 & \second{66.93} & \second{44.69} & \second{39.60} \\
\rowcolor{faithblue}
FAITH (DWT) & \best{78.92} & \best{64.13} & \best{61.91} & \best{76.92} & \best{62.75} & \best{59.86} & \best{67.19} & \best{44.79} & \best{39.71} \\
\bottomrule
\end{tabular}
\end{adjustbox}
\end{subtable}

\vspace{2mm}
\scriptsize
\begin{subtable}[t]{\linewidth}
\centering
\caption{}
\vspace{-.1in}\small
\label{tab:noise}
\begin{adjustbox}{max width=0.98\linewidth,center}
\begin{tabular}{l *{3}{ccc}}
\toprule
\multirow{2}{*}{\textbf{Methods}} &
\multicolumn{3}{c}{Noise intensity -- 10\%} &
\multicolumn{3}{c}{Noise intensity -- 15\%} &
\multicolumn{3}{c}{Noise intensity -- 20\%} \\
\cmidrule(lr){2-4} \cmidrule(lr){5-7} \cmidrule(lr){8-10}
& \textit{Fixed} & \textit{Adaptive} & \textit{Full} &
\textit{Fixed} & \textit{Adaptive} & \textit{Full} &
\textit{Fixed} & \textit{Adaptive} & \textit{Full} \\
\midrule
Shuai \textit{et al.}~\cite{shuai2023locateverify} & 49.23 & 29.46 & 25.66 & 43.09 & 18.97 & 15.05 & 40.72 & 15.81 & 8.12 \\
FreqNet~\cite{tan2024freqnet}                        & 48.88 & 29.03 & 25.38 & 42.84 & 18.96 & 15.78 & 40.36 & 15.82 & 8.26 \\
Ba \textit{et al.}~\cite{ba2024exposingdeception}
                                              & 49.32 & 30.03 & 25.76 & 42.98 & 18.99 & 15.80 & 40.53 & 16.22 & 8.49 \\
SeqFakeFormer~\cite{shao2022sequential}
                                              & 62.39 & 34.59 & 31.62 & 55.61 & 23.40 & 19.28 & 51.21 & 20.27 & 11.31 \\
\midrule
\rowcolor{faithblue}
FAITH (DCT) & 62.68 & \best{35.52} & \best{32.16} & \second{55.71} & \second{23.59} & \second{19.48} & 51.32 & 20.26 & 11.49 \\
\rowcolor{faithblue}
FAITH (FFT) & \second{62.91} & 34.37 & 31.76 & 55.28 & 23.06 & 18.95 & \second{51.51} & \second{20.39} & \second{11.83} \\
\rowcolor{faithblue}
FAITH (DWT) & \best{63.07} & \second{34.69} & \second{31.88} & \best{55.84} & \best{24.38} & \best{19.60} & \best{51.70} & \best{20.83} & \best{12.19} \\
\bottomrule
\end{tabular}
\end{adjustbox}
\end{subtable}
\end{table*}

\subsection{Robustness Study}
In realistic image transmission and storage, post-processing operations can distort provenance cues.
We therefore evaluate robustness under JPEG compression and additive Gaussian noise, as summarized in Table~\ref{tab:robustness}.

\noindent\textbf{JPEG compression.}
Table~\ref{tab:jpeg} shows that all methods degrade as compression becomes stronger, and the drop is most pronounced at 75\% compression.
This is consistent with JPEG removing high-frequency details through quantization, which can weaken both spatial textures and frequency-domain signals.
Even under these conditions, FAITH variants remain strong relative to baselines and maintain higher Full-Acc across compression levels.
Among the three transforms, FAITH(DWT) is consistently best or second-best in most columns, suggesting that its multi-resolution decomposition retains useful cues even when fine details are partially suppressed.

\noindent\textbf{Gaussian noise.}
Table~\ref{tab:noise} reports results under increasing noise intensity.
Unlike compression, additive noise perturbs nearly all frequencies and directly corrupts both edges and textures, so degradation is visible for every method.
Despite this, FAITH(DWT) attains the best Full-Acc at higher noise levels and remains comparatively stable relative to baselines, indicating that directional high-frequency representations still provide discriminative signals under moderate stochastic corruption.
At the same time, performance under severe noise remains low for all models, which highlights an open robustness gap for sequential provenance tracing in uncontrolled environments.


\definecolor{HLgreen}{RGB}{232,245,233}  
\definecolor{HLred}{RGB}{255,235,238}    
\definecolor{GTolive}{RGB}{85,107,47}    

\newcommand{\EOS}{\textcolor{gray}{EOS}}
\newcommand{\arr}{\,\(\to\)\,}

\newcommand{\hlg}[1]{\begingroup\setlength{\fboxsep}{0.6pt}\colorbox{HLgreen}{\strut #1}\endgroup}
\newcommand{\hlr}[1]{\begingroup\setlength{\fboxsep}{0.6pt}\colorbox{HLred}{\strut #1}\endgroup}
\newcommand{\gtc}[1]{\textcolor{GTolive}{#1}}

\newcommand{\imgborder}[1]{%
  \begingroup
  \setlength{\fboxsep}{0pt}%
  \setlength{\fboxrule}{0.6pt}%
  \fcolorbox{black}{white}{#1}%
  \endgroup
}

\begin{figure*}[t!]
\centering
\fontsize{1.}{1.}\selectfont

\setlength{\tabcolsep}{4pt}
\renewcommand{\arraystretch}{1.05}

\begin{minipage}[t]{\textwidth}
\begin{minipage}[t]{0.28\textwidth}
  \vspace{0pt}
  \centering
  \imgborder{\includegraphics[width=\linewidth]{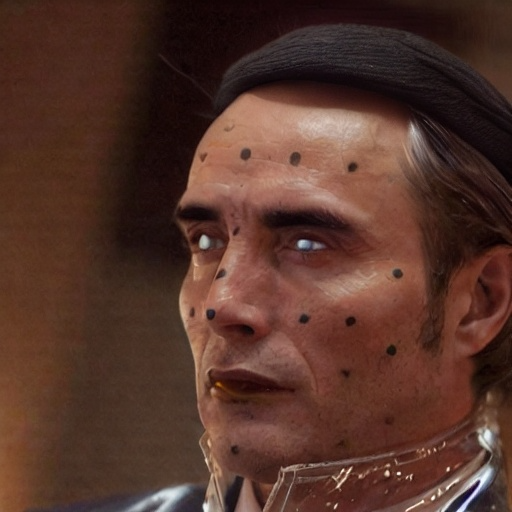}}
\end{minipage}\hfill
\begin{minipage}[t]{0.70\textwidth}
  \vspace{0pt}
  \resizebox{\linewidth}{!}{%
  \begin{tabular}{@{}l p{0.80\linewidth}@{}}
    \toprule
    \multicolumn{2}{@{}l@{}}{\textbf{Clean}} \\
    \cmidrule(lr){1-2}
    \textbf{\gtc{GT}} & \gtc{Lipstick}\arr\gtc{Hat}\arr\gtc{Eye}\arr\gtc{Eyebrow} \\
    \textbf{Ours} & \hlg{Lipstick}\arr\hlg{Hat}\arr\hlg{Eye}\arr\hlg{Eyebrow} \\
    \textbf{SeqFakeFormer} & \hlg{Lipstick}\arr\hlg{Hat}\arr\hlg{Eye}\arr\hlg{Eyebrow} \\
    \textbf{FreqNet} & \hlg{Lipstick}\arr\hlr{Eye}\arr\hlr{\EOS} \\
    \midrule

    \multicolumn{2}{@{}l@{}}{\textbf{Compress-50\%}} \\
    \cmidrule(lr){1-2}
    \textbf{\gtc{GT}} & \gtc{Lipstick}\arr\gtc{Hat}\arr\gtc{Eye}\arr\gtc{Eyebrow} \\
    \textbf{Ours} & \hlg{Lipstick}\arr\hlg{Hat}\arr\hlg{Eye}\arr\hlg{Eyebrow} \\
    \textbf{SeqFakeFormer} & \hlg{Lipstick}\arr\hlg{Hat}\arr\hlr{\EOS} \\
    \textbf{FreqNet} & \hlg{Lipstick}\arr\hlr{\EOS} \\
    \midrule

    \multicolumn{2}{@{}l@{}}{\textbf{Noise intensity-15\%}} \\
    \cmidrule(lr){1-2}
    \textbf{\gtc{GT}} & \gtc{Lipstick}\arr\gtc{Hat}\arr\gtc{Eye}\arr\gtc{Eyebrow} \\
    \textbf{Ours} & \hlg{Lipstick}\arr\hlg{Hat}\arr\hlg{Eye}\arr\hlg{Eyebrow} \\
    \textbf{SeqFakeFormer} & \hlg{Lipstick}\arr\hlg{Hat}\arr\hlg{Eye}\arr\hlr{\EOS} \\
    \textbf{FreqNet} & \hlg{Lipstick}\arr\hlr{\EOS} \\
    \bottomrule
  \end{tabular}}
\end{minipage}
\end{minipage}

\vspace{2.0mm}

\begin{minipage}[t]{\textwidth}
\begin{minipage}[t]{0.28\textwidth}
  \vspace{0pt}
  \centering
  \imgborder{\includegraphics[width=\linewidth]{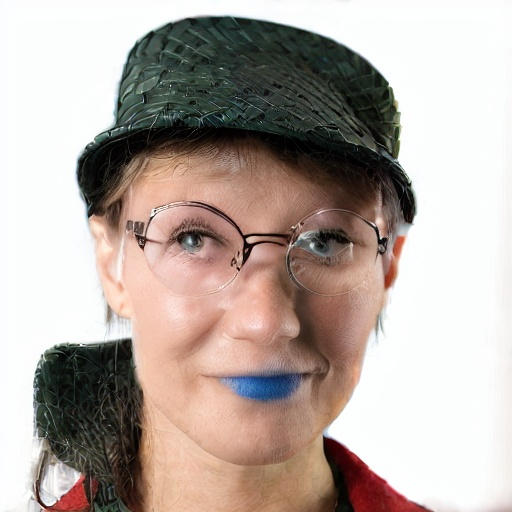}}
\end{minipage}\hfill
\begin{minipage}[t]{0.70\textwidth}
  \vspace{0pt}
  \resizebox{\linewidth}{!}{%
  \begin{tabular}{@{}l p{0.80\linewidth}@{}}
    \toprule
    \multicolumn{2}{@{}l@{}}{\textbf{Clean}} \\
    \cmidrule(lr){1-2}
    \textbf{\gtc{GT}} & \gtc{Glasses}\arr\gtc{Eyebrow}\arr\gtc{Lipstick}\arr\gtc{Hat} \\
    \textbf{Ours} & \hlg{Glasses}\arr\hlg{Eyebrow}\arr\hlg{Lipstick}\arr\hlg{Hat} \\
    \textbf{SeqFakeFormer} & \hlg{Glasses}\arr\hlr{Hat}\arr\hlr{Eyebrow}\arr\hlr{Lipstick} \\
    \textbf{FreqNet} & \hlg{Glasses}\arr\hlr{Hat}\arr\hlr{\EOS} \\
    \midrule

    \multicolumn{2}{@{}l@{}}{\textbf{Compress-50\%}} \\
    \cmidrule(lr){1-2}
    \textbf{\gtc{GT}} & \gtc{Glasses}\arr\gtc{Eyebrow}\arr\gtc{Lipstick}\arr\gtc{Hat} \\
    \textbf{Ours} & \hlg{Glasses}\arr\hlg{Eyebrow}\arr\hlg{Lipstick}\arr\hlg{Hat} \\
    \textbf{SeqFakeFormer} & \hlg{Glasses}\arr\hlr{Hat}\arr\hlg{Lipstick}\arr\hlr{\EOS} \\
    \textbf{FreqNet} & \hlg{Glasses}\arr\hlr{Hat}\arr\hlr{\EOS} \\
    \midrule

    \multicolumn{2}{@{}l@{}}{\textbf{Noise intensity-15\%}} \\
    \cmidrule(lr){1-2}
    \textbf{\gtc{GT}} & \gtc{Glasses}\arr\gtc{Eyebrow}\arr\gtc{Lipstick}\arr\gtc{Hat} \\
    \textbf{Ours} & \hlg{Glasses}\arr\hlg{Eyebrow}\arr\hlg{Lipstick}\arr\hlg{Hat} \\
    \textbf{SeqFakeFormer} & \hlg{Glasses}\arr\hlg{Eyebrow}\arr\hlr{\EOS} \\
    \textbf{FreqNet} & \hlg{Glasses}\arr\hlr{Hat}\arr\hlr{\EOS} \\
    \bottomrule
  \end{tabular}}
\end{minipage}
\end{minipage}

\caption{Qualitative robustness examples under Compress-50\% and Noise intensity-15\%. Tokens are replaced with semantic attributes. \EOS\ denotes early stopping. Correct tokens are highlighted in light green and incorrect tokens in light red. Ground truth is shown in olive-green text.}
\label{fig:qual_robust}
\vspace{-5mm}
\end{figure*}

\subsection{Qualitative Robustness Analysis under Compression and Noise}
\label{subsec:qual_robust}

To complement the quantitative robustness results, we present representative qualitative cases under two common post-processing perturbations, i.e., \emph{Compress-50\%} and \emph{Noise intensity-15\%} (Fig.~\ref{fig:qual_robust}). We replace token IDs with semantic facial attributes and highlight position-wise correctness: green indicates a token matching the ground-truth attribute at the same step, while red denotes an incorrect token or early stopping (\texttt{EOS}). This visualization reveals \emph{how} perturbations affect sequential provenance tracing beyond aggregate accuracy.

Across both examples, our method preserves the complete manipulation trajectory under clean as well as degraded inputs, consistently recovering the full four-step sequences. In contrast, competing baselines often exhibit two dominant failure modes under compression or noise: (1) \textbf{premature termination} (early \texttt{EOS}), leading to missing later edits, and (2) \textbf{step-wise drift} where the predicted attribute at a given step shifts or swaps the intended order. Notably, under \emph{Compress-50\%}, SeqFakeFormer frequently collapses to shorter sequences, while FreqNet tends to terminate even earlier, suggesting that their evidence is less stable once high-frequency cues are attenuated. Under \emph{Noise intensity-15\%}, the baselines further suffer from occasional step mismatches, indicating increased uncertainty in deciding both \emph{what} was edited and \emph{when} it occurred.

These observations support the key advantage of our frequency-aware design: by leveraging frequency-domain guidance to stabilize cross-step evidence, the model remains less sensitive to post-processing artifacts that obscure subtle traces. As a result, our predictions are not only more accurate, but also more \emph{complete} (avoiding early stopping) and more \emph{order-consistent} (maintaining correct step-wise alignment). This qualitative analysis aligns with the robustness tables and suggests that frequency-aware cues are particularly beneficial for sequential provenance tracing where each step may leave weak, partially overwritten footprints that are easily disrupted by compression and noise.

\section{Conclusion}
We introduce SEED, the first large-scale benchmark for sequential visual provenance tracing in facial imagery, with over 90,000 diffusion-edited images and fine-grained metadata (edit order and prompts). SEED models ordered edit trajectories and provides a sequence-length-balanced split and sequence-centric metrics that separate token-level correctness from strict history recovery. Our evaluation shows that single-step manipulation detectors transfer poorly to sequential tracing and that performance degrades as chain length increases, highlighting the challenge of compositional and partially overlapping edits. To support benchmarking, we further introduce FAITH, a frequency-aware Transformer baseline that integrates spatial and high-frequency signals. FAITH provides stable improvements across sequence lengths and under common post-processing perturbations, with wavelet components yielding the most consistent gains.

\bibliographystyle{splncs04}
\bibliography{main}

\appendix
\newcommand{\tokbox}[2]{%
  \begingroup
  \setlength{\fboxsep}{1pt}%
  \colorbox{#1}{\strut #2}%
  \endgroup
}

\newcommand{\tokok}[1]{\tokbox{green!20}{#1}}
\newcommand{\tokbad}[1]{\tokbox{red!20}{#1}}
\newcommand{\tokpad}[1]{\tokbox{gray!20}{#1}}
\renewcommand{\thesection}{\Alph{section}}

\title{Supplementary Material for\\
\textit{SEED: A Large-Scale Benchmark for Provenance Tracing in Sequential Deepfake Facial Edits}}
\author{}        
\institute{}     
\titlerunning{Supplementary Material}
\maketitle

\beginsupp

\section{Supplementary Dataset Protocol Details}
Table~\ref{tab:forgery_datasets_supp_protocol} complements the dataset comparison table in the main paper by documenting protocol-level details that are difficult to communicate through a compact checkmark matrix.
Across deepfake detection and synthetic image forensics, recent surveys consistently argue that reported generalization and robustness are highly contingent on dataset construction and evaluation design, including the choice of generation/editing pipelines, the presence of standardized perturbation suites, the extent of in-the-wild post-processing, and whether the intended evaluation emphasizes cross-generator or cross-domain transfer~\cite{wang2024deepfakedetection,liu2024evolvingmultimodal,nguyenle2024passive,xie2025aigc}.
In other words, superficially similar ``detection'' settings can correspond to different threat models and difficulty factors once the underlying protocol is made explicit.

Protocol-oriented reviews of AI-generated image detection further highlight this issue by organizing widely used benchmarks according to their generation sources, annotation availability, and evaluation setups, and by discussing representative datasets such as GenImage and ImagiNet in the context of diffusion-era detection cues and robustness testing~\cite{mahara2025generated,xie2025aigc}.
Motivated by these observations, Table~\ref{tab:forgery_datasets_supp_protocol} summarizes for each dataset (i) the major generation/editing pipelines and data sources, (ii) dataset-provided supervision (e.g., order, attribute labels, prompts, and spatial annotations when available), and (iii) the dataset-defined evaluation protocol, including perturbations and any prescribed cross-generator or cross-domain splits.
By making these protocol choices explicit, Table~S1 enables readers to interpret cross-dataset comparisons under a consistent understanding of what each benchmark is designed to test and why protocol differences materially affect conclusions.

\begin{table}[t!]
\centering
\caption{Protocol-level details for image-based deepfake benchmarks.
We summarize major editing/generation methods and the dataset-specific evaluation or perturbation protocols.}
\label{tab:forgery_datasets_supp_protocol}

\tiny
\setlength{\tabcolsep}{3.2pt}
\renewcommand{\arraystretch}{1.22}

\begingroup
\hyphenpenalty=10000
\exhyphenpenalty=10000
\sloppy
\setlength{\emergencystretch}{2em}

\begin{tabularx}{\columnwidth}{@{} L{2.75cm} Y Y @{}}
\toprule
\rowcolor{tblHead}
\textbf{Dataset} &
\textbf{Forgery / Editing Methods} &
\textbf{Annotations, Tasks, and Protocol (incl. perturbations if defined)} \\
\midrule

\SectRowS{A. Image Detection / Localization}

\rowcolor{tblLite}
OpenForensics~\cite{le2021openforensics} &
In-the-wild multi-face forgery benchmark with diverse manipulation sources; emphasizes realistic, unconstrained scenarios. &
Provides face-wise spatial supervision for localization (e.g., masks/boundaries) in addition to detection; includes challenge settings with difficulty-aware evaluation and augmentation-based configurations (not a single fixed perturbation suite). \\

\rowcolor{tblLite}
ForgeryNet~\cite{he2021forgerynet} &
Mega-scale benchmark with heterogeneous synthesis/editing pipelines; covers identity-replaced and identity-remained manipulations (we use the image subset in our comparison). &
Supports detection and (for many subsets) localization with rich metadata; protocol is dataset-driven rather than a single mandatory suite, and is commonly evaluated under cross-domain and distortion robustness settings. \\

\rowcolor{tblLite}
DeepFakeFace-DFF~\cite{song2023deepfakeface} &
Diffusion-centric fake face generation (e.g., Stable Diffusion v1.5 and inpainting-style generation), plus toolbox-based synthesis variants. &
Primarily designed for diffusion-era detection; generation-focused setup and typically used without a standardized perturbation suite defined by the dataset. \\

\rowcolor{tblLite}
DiffusionFace~\cite{chen2024diffusionface} &
Diffusion-based face forgery dataset spanning multiple categories (e.g., T2I, I2I, inpainting, and swap-style settings) built from multiple diffusion models. &
Provides generator/category metadata to support diffusion forensics; focuses on generation diversity, and protocols are typically cross-generator generalization tests rather than a single unified perturbation suite. \\

\rowcolor{tblLite}
GenImage~\cite{zhu2023genimage} &
Large-scale benchmark for synthetic image detection across multiple generators (GANs and diffusion models), designed for generalizable detection. &
Evaluation emphasizes cross-generator and cross-domain generalization; commonly reports robustness under standard degradations such as JPEG compression, resizing, and blur (dataset provides standard splits and guidance). \\

\rowcolor{tblLite}
DRCT-2M~\cite{chen2024drct} &
Large-scale diffusion-generated image detection benchmark covering multiple diffusion model families and settings. &
Evaluation-oriented benchmark highlighting generalization across diffusion models and data sources; commonly used with cross-model testing, without enforcing a single fixed perturbation suite. \\

\rowcolor{tblLite}
ImagiNet~\cite{boychev2024imaginet} &
High-resolution benchmark with matched real and synthetic images across diverse content types (including faces); includes multiple generators (as reported). &
Supports detection (real vs. synthetic) and, in some configurations, generator attribution; evaluation often reports robustness under common post-processing such as resizing/compression. \\

\rowcolor{tblLite}
Fake2M~\cite{lu2023seeing} &
Large-scale synthetic image benchmark with millions of AI-generated images paired with real photographs, used for detection and perception studies. &
Protocol emphasizes large-scale evaluation and, in some uses, human perceptual analysis; not restricted to a single fixed perturbation suite, with perturbations often applied as evaluation-time stress tests. \\

\addlinespace[0.25em]
\midrule

\SectRowS{B. Provenance Tracing (Sequential)}

\rowcolor{tblLite}
Seq-DeepFake (components)~\cite{shao2022sequential} &
GAN-based sequential \emph{component} manipulation (e.g., transplanting facial parts) with multi-step trajectories. &
Provides step-wise edit order supervision and component-level labels for sequence prediction (provenance tracing); the base dataset does not define a standardized perturbation protocol. \\

\rowcolor{tblLite}
Seq-DeepFake (attributes)~\cite{shao2022sequential} &
GAN-based sequential \emph{attribute} editing using StyleGAN-style editing pipelines (e.g., Talk-to-Edit), forming multi-step trajectories. &
Provides step-wise edit order supervision and attribute labels for sequence prediction; the base dataset does not define a standardized perturbation protocol. \\

\rowcolor{tblLite}
Seq-DeepFake-P~\cite{shao2025robust} &
Robustness extension of Seq-DeepFake (components + attributes) to stress-test sequential detection/tracing. &
Defines a standardized perturbation suite: 6 perturbations $\times$ 3 levels plus mixed settings (e.g., JPEG, WGN, blur, saturation, contrast, block-wise distortion). \\

\rowcolor{seedHL}
\textbf{SEED (Ours)} &
\textbf{Diffusion-based sequential, localized, instruction-driven edits with per-step editor sampling (e.g., LEdits, SDXL, SD3-based pipelines), producing 1--4 step trajectories.} &
\textbf{Records step-wise provenance (attribute, prompt, mask, editor) and quality measurements as metadata, enabling detection, localization, and edit-order recovery; dataset release focuses on provenance logging rather than enforcing a single fixed perturbation suite (perturbations are applied consistently at evaluation time).} \\

\bottomrule
\end{tabularx}

\vspace{0.2em}
{\tiny \textit{Remark.} We omit compact columns (e.g., Mod./Scale/Seq/Sup.) to allocate space to protocol descriptions; the omitted fields are reported in the main paper.}
\endgroup
\end{table}

\section{SEED Dataset Completeness and Utility}
\label{sec:seed_completeness}

\subsection{Step-wise Quality Measurements}
\label{subsec:seed_quality}

For each edit step $t$, we log four complementary quality measurements:
(i) \textbf{CLIP-Sim}, text-image alignment between the instantiated prompt and the edited image;
(ii) \textbf{DINOv2-Sim}, semantic similarity between the pre-edit and post-edit images in DINOv2 feature space;
(iii) \textbf{SSIM}, structural similarity between the pre-edit and post-edit images;
and (iv) \textbf{CLIP-Image-Sim}, image-image similarity in CLIP space.
These scores are recorded for auditing dataset difficulty and for downstream analysis of edit reliability and trace preservation, rather than enforcing narrow acceptance bands.

Table~\ref{tab:seed_step_quality_no_n} reports step-wise statistics grouped by edit index $t\in\{1,2,3,4\}$, including the mean, standard deviation, and percentile summary (P25/P50/P95).
Overall, image-image similarity metrics (DINOv2/SSIM/CLIP-Image-Sim) remain high across steps, while CLIP-Sim exhibits a mild decrease as the edit chain grows longer, consistent with later edits being applied on an already edited face where textual grounding can be less deterministic.

\begin{table}[t]
\centering
\small
\setlength{\tabcolsep}{5.2pt}
\renewcommand{\arraystretch}{1.15}
\caption{\textbf{Step-wise quality statistics in SEED (without counts).}
We report mean, Std, and percentiles (P25/P50/P95) for each metric at edit index $t$.
All values are rounded to two decimals.}
\label{tab:seed_step_quality_no_n}
\begin{tabular}{c l c c c c c}
\toprule
\textbf{$t$} & \textbf{Metric} & \textbf{Mean} & \textbf{Std} & \textbf{P25} & \textbf{P50} & \textbf{P95} \\
\midrule
\multirow{4}{*}{1}
 & CLIP-Sim       & 0.23 & 0.04 & 0.20 & 0.23 & 0.29 \\
 & DINOv2-Sim     & 0.94 & 0.07 & 0.93 & 0.97 & 1.00 \\
 & SSIM           & 0.85 & 0.07 & 0.80 & 0.85 & 0.95 \\
 & CLIP-Image-Sim & 0.91 & 0.07 & 0.86 & 0.92 & 0.99 \\
\midrule
\multirow{4}{*}{2}
 & CLIP-Sim       & 0.22 & 0.04 & 0.20 & 0.22 & 0.29 \\
 & DINOv2-Sim     & 0.95 & 0.07 & 0.94 & 0.98 & 1.00 \\
 & SSIM           & 0.89 & 0.07 & 0.85 & 0.91 & 0.97 \\
 & CLIP-Image-Sim & 0.92 & 0.06 & 0.89 & 0.94 & 0.99 \\
\midrule
\multirow{4}{*}{3}
 & CLIP-Sim       & 0.22 & 0.03 & 0.19 & 0.22 & 0.28 \\
 & DINOv2-Sim     & 0.95 & 0.07 & 0.95 & 0.98 & 1.00 \\
 & SSIM           & 0.90 & 0.06 & 0.87 & 0.92 & 0.97 \\
 & CLIP-Image-Sim & 0.93 & 0.05 & 0.91 & 0.95 & 0.99 \\
\midrule
\multirow{4}{*}{4}
 & CLIP-Sim       & 0.22 & 0.03 & 0.19 & 0.21 & 0.28 \\
 & DINOv2-Sim     & 0.96 & 0.05 & 0.96 & 0.99 & 1.00 \\
 & SSIM           & 0.91 & 0.06 & 0.88 & 0.93 & 0.98 \\
 & CLIP-Image-Sim & 0.94 & 0.04 & 0.92 & 0.95 & 0.99 \\
\bottomrule
\end{tabular}
\end{table}

\subsection{Metadata Schema and File Organization}
\label{subsec:seed_schema}

SEED is released with per-sample JSONL metadata, where the first line stores trajectory-level information and each subsequent line corresponds to one edit step.
This design enables faithful reconstruction of the full provenance chain and supports evaluation for trajectory prediction and spatial evidence analysis.

Table~\ref{tab:seed_schema} summarizes the metadata schema.
We use \texttt{file\_name} to index samples, \texttt{sequence} to store the ordered attribute chain, and \texttt{model\_sequence} to record the diffusion editor identifier at each step.
Each step line includes the attribute label (\texttt{seq}), editor index (\texttt{model\_idx}), prompt/caption, mask information, generation hyperparameters (e.g., \texttt{num\_inversion\_steps}, \texttt{guidance\_scale}, \texttt{image\_guidance\_scale}), and step-wise metrics.
Some fields such as \texttt{seed} may be present for specific editors.

\noindent\textbf{JSONL example.}
{
\scriptsize{
\begin{verbatim}

{"file_name":"99930","sequence":[1,2,6,3],"model_sequence":[2,2,2,1],
 "prompts":["Change her eyes color to blue.","Turn her lipstick color to blue.",
           "Make her eyebrow thick.","A girl with purple hair."]}
{"seq":1,"model_idx":2,"num_inversion_steps":60,"guidance_scale":8,"image_guidance_scale":2,
 "prompt":"Change her eyes color to blue.","caption":"A girl with blue eyes.",
 "mask_exists":false,"mask_path":"./mask/12.png",
 "metrics":{"clip_sim":0.2639,"clip_image_sim":0.8182,"dinov2_sim":0.9937,"ssim":0.9549}}
\end{verbatim}
}
}

\begin{table}[t]
\centering
\small
\setlength{\tabcolsep}{4.8pt}
\renewcommand{\arraystretch}{1.15}
\caption{\textbf{SEED metadata schema (JSONL).}
The first line is trajectory-level metadata; each subsequent line corresponds to one edit step.}
\label{tab:seed_schema}
\begin{adjustbox}{max width=\linewidth,center}
\begin{tabular}{l l p{0.62\linewidth}}
\toprule
\textbf{Level} & \textbf{Field} & \textbf{Description} \\
\midrule
Trajectory
& \texttt{file\_name} & Sample identifier (used to locate images and masks). \\
& \texttt{sequence} & Ordered attribute IDs, length $L\in\{1,2,3,4\}$. \\
& \texttt{model\_sequence} & Ordered editor IDs aligned with \texttt{sequence}. \\
& \texttt{prompts} & Instantiated prompts used at each step (optional redundancy for convenience). \\
\midrule
Step
& \texttt{seq} & Attribute ID of the current step. \\
& \texttt{model\_idx} & Editor ID used at this step. \\
& \texttt{num\_inversion\_steps} & Inversion or editing steps for the diffusion editor. \\
& \texttt{guidance\_scale} & Classifier-free guidance scale (editor-dependent). \\
& \texttt{image\_guidance\_scale} & Image guidance scale (editor-dependent). \\
& \texttt{seed} & Random seed (optional; present for some editors). \\
& \texttt{prompt} / \texttt{caption} & Instruction-style prompt and its caption-style form. \\
& \texttt{mask\_exists} & Whether a valid spatial mask is used. \\
& \texttt{mask\_path} & Path to mask file (typically relative to the sample directory). \\
& \texttt{metrics} & Step-wise scores: \texttt{clip\_sim}, \texttt{clip\_image\_sim}, \texttt{dinov2\_sim}, \texttt{ssim}. \\
\bottomrule
\end{tabular}
\end{adjustbox}
\end{table}

\subsection{Split Protocol}
\label{subsec:seed_split}

SEED uses an identity-disjoint split for training, validation, and testing.
Specifically, identities do not overlap across splits, preventing identity leakage and ensuring that reported results are not inflated by subject memorization.
Since the split is identity-based by design, we do not additionally report a separate ``cross-identity generalization'' setting beyond the standard held-out test split.

\begin{figure}[t]
  \centering
  \includegraphics[width=0.7\linewidth]{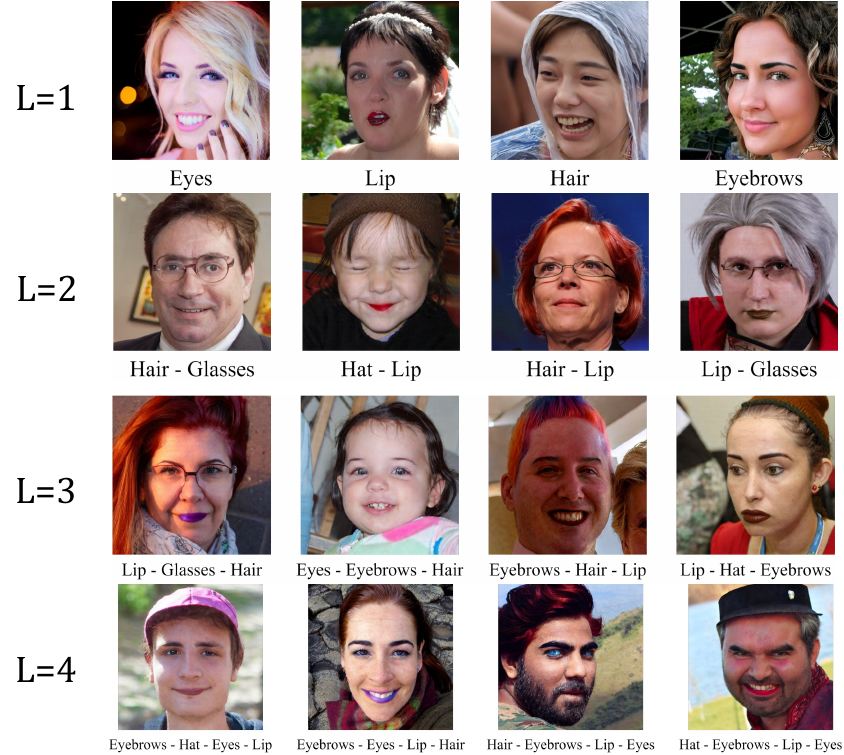}
  \caption{\textbf{Representative SEED trajectories with increasing sequence length.}
  Each group corresponds to a sequence length $L\in\{1,2,3,4\}$.
  Labels under each example denote the ordered edit chain (e.g., \textit{Hair-Glasses} or \textit{Eyebrows-Hat-Eyes-Lip}), illustrating compositional attribute edits and the growing difficulty of recovering complete provenance histories as the edit chain becomes longer.}
  \label{fig:seed_samples}
\end{figure}

Figure~\ref{fig:seed_samples} visualizes the core challenge targeted by SEED.
For $L=1$, evidence is relatively localized and often dominated by a single semantic change.
As $L$ increases to 2--4, edits become compositional (intrinsic attributes and accessories), and later operations can suppress earlier traces, producing realistic images whose provenance history is harder to reconstruct from the final output alone.
These examples motivate benchmarks with step-wise provenance supervision and analysis methods that can aggregate weak evidence across space to recover both attribute identity and temporal order under diffusion-based sequential editing.

\section{Experimental Details}
This supplementary section provides additional experimental specifications to facilitate reproducibility and clarify evaluation choices summarized in the main paper.
In particular, we detail the sequence-level metrics used for edit-trace recovery, report the implementation configuration of FAITH/SeqFakeFormer, and provide runtime efficiency measurements and error-mode analysis to contextualize model behavior under increasing edit-chain length.

\subsection{Evaluation Metrics}
\noindent\textbf{Motivation and sources.}
Sequential provenance tracing is naturally formulated as an image-to-sequence prediction problem, where a model outputs an ordered chain of manipulated attributes, consistent with prior sequential deepfake manipulation settings~\cite{shao2022sequential,shao2025robust}.
Following standard practice in sequence prediction, we report (i) token-level accuracy under a fixed horizon (to reflect per-step correctness), (ii) strict exact-match accuracy (to reflect complete history recovery), and (iii) an overlap-only variant that reduces the dominance of padded positions when sequences are shorter than the maximum length.

\vspace{0.25em}
\noindent\textbf{Notation.}
Let the ground-truth attribute sequence be $\mathbf{y}=(y_1,\ldots,y_T)$ with length $T=|\mathbf{y}|$, and the predicted sequence be $\hat{\mathbf{y}}=(\hat{y}_1,\ldots,\hat{y}_{\hat{T}})$ with $\hat{T}=|\hat{\mathbf{y}}|$.
We set the maximum evaluated length to $T_{\max}=4$ (the SEED protocol).
Let $\varnothing$ denote the padding token \texttt{no-manipulation}.
Define the padding operator $\mathrm{Pad}(\cdot)$ that maps any sequence to length $T_{\max}$:
\begin{equation}
\tilde{\mathbf{y}} = \mathrm{Pad}(\mathbf{y}) \in (\mathcal{V}\cup\{\varnothing\})^{T_{\max}},\quad
\tilde{y}_i =
\begin{cases}
y_i, & i \le \min(T,T_{\max}),\\
\varnothing, & \text{otherwise},
\end{cases}
\label{eq:pad_def}
\end{equation}
and similarly $\tilde{\hat{\mathbf{y}}}=\mathrm{Pad}(\hat{\mathbf{y}})$.

\vspace{0.15em}
\noindent\textbf{Fixed-Acc (position-wise accuracy under fixed length).}
We compare the padded sequences at all $T_{\max}$ positions:
\begin{equation}
\mathrm{Fixed\mbox{-}Acc}(\hat{\mathbf{y}},\mathbf{y})
=
\frac{1}{T_{\max}}
\sum_{i=1}^{T_{\max}}
\mathbb{I}\!\left[\tilde{\hat{y}}_i=\tilde{y}_i\right].
\label{eq:fixed_acc}
\end{equation}
This metric penalizes both wrong attributes and wrong sequence lengths, because missing edits become mismatches against non-padding targets, and extra edits become mismatches against $\varnothing$.

\vspace{0.15em}
\noindent\textbf{Adaptive-Acc (overlap-only prefix accuracy).}
To avoid overweighting padded positions, we compare only the overlapping prefix up to $m=\min(\hat{T},T)$:
\begin{equation}
\mathrm{Adaptive\mbox{-}Acc}(\hat{\mathbf{y}},\mathbf{y})
=
\begin{cases}
1, & m=0 \ \text{and}\ \hat{T}=T=0,\\
0, & m=0 \ \text{and}\ (\hat{T}\neq 0 \ \text{or}\ T\neq 0),\\
\frac{1}{m}\sum_{i=1}^{m}\mathbb{I}\!\left[\hat{y}_i=y_i\right], & m>0.
\end{cases}
\label{eq:adaptive_acc}
\end{equation}
Thus Adaptive-Acc focuses on correctness where both sequences overlap, and is less sensitive to trailing missing or extra steps.

\vspace{0.15em}
\noindent\textbf{Full-Acc (strict exact match).}
We count a sample as correct only when the entire predicted sequence exactly matches the ground-truth in both length and content:
\begin{equation}
\mathrm{Full\mbox{-}Acc}(\hat{\mathbf{y}},\mathbf{y})
=
\mathbb{I}\!\left[\hat{T}=T\right]\cdot
\prod_{i=1}^{T}
\mathbb{I}\!\left[\hat{y}_i=y_i\right].
\label{eq:full_acc}
\end{equation}
This is the strictest metric and directly measures complete edit-history recovery.

\vspace{0.35em}
\noindent\textbf{Single illustrative example.}
Let $T_{\max}=4$, ground-truth $\mathbf{y}=[\texttt{Lip},\texttt{Hat},\texttt{Eyes}]$, and prediction $\hat{\mathbf{y}}=[\texttt{Lip},\texttt{Hat}]$ (an under-prediction case).
After padding:
\begin{center}
\begin{tabular}{@{}l@{}}
\textbf{GT:} [\tokok{Lip}, \tokok{Hat}, \tokbad{Eyes}, \tokpad{no-manipulation}]\\
\textbf{PR:} [\tokok{Lip}, \tokok{Hat}, \tokpad{no-manipulation}, \tokpad{no-manipulation}]
\end{tabular}
\end{center}
\noindent\textbf{Fixed-Acc:} $\mathrm{Fixed\mbox{-}Acc}=3/4$ by Eq.~\eqref{eq:fixed_acc}. \quad
\textbf{Adaptive-Acc:} $\mathrm{Adaptive\mbox{-}Acc}=2/2$ by Eq.~\eqref{eq:adaptive_acc}. \quad
\textbf{Full-Acc:} $\mathrm{Full\mbox{-}Acc}=0$ by Eq.~\eqref{eq:full_acc}.

\vspace{0.25em}
\noindent\textbf{Interpretation.}
Fixed-Acc reflects both attribute correctness and length errors under a fixed horizon, Adaptive-Acc isolates correctness on the overlapping portion (reducing padding bias), and Full-Acc enforces strict recovery of the complete edit history, penalizing any mistake in attribute identity, order, or length.

\subsection{Implementation Configuration}
Table~\ref{tab:faith_impl_config} lists the configuration used for FAITH/SeqFakeFormer, including the backbone, transformer capacity, and optimization schedule.

\begin{table}[t]
\centering
\scriptsize
\setlength{\tabcolsep}{5pt}
\renewcommand{\arraystretch}{1.12}
\caption{\textbf{Implementation configuration of FAITH/SeqFakeFormer.}}
\label{tab:faith_impl_config}
\begin{tabular}{l l}
\toprule
\textbf{Component} & \textbf{Setting} \\
\midrule
Backbone &
ResNet-50; sine positional embedding; FrozenBN: False \\
\midrule
Input &
Image size: $512 \times 512$ \\
\midrule
Transformer &
Hidden dim: 512; FFN dim: 512; heads: 4; dropout: 0.1; pre-norm: True \\
&
Encoder layers: 2; Decoder layers: 2 \\
&
Vocab size: 9; max position embeddings: 5 (output length $L=4$) \\
&
Tokens: SOS=0, EOS=7, PAD=8 \\
\midrule
Optimization &
Batch size: 48; epochs: 150; weight decay: $10^{-4}$; grad clip: 0.1 \\
&
LR (backbone): $10^{-4}$; LR (others): $10^{-3}$ \\
&
Warm-up: 25 epochs; LR milestones: [50, 75, 100] \\
\bottomrule
\end{tabular}
\end{table}

\subsection{Runtime Efficiency and Throughput}
To quantify inference cost, we benchmark end-to-end decoding throughput under the same autoregressive protocol used in evaluation (maximum length $4$ with early stopping by EOS).
We report parameter count, throughput (images per second), latency (milliseconds per image), and GPU memory usage.
The benchmark uses batch size 16 with AMP enabled, a 10-batch warm-up, and times inference over 512 images.
Peak memory is measured using \texttt{torch.cuda.max\_memory\_allocated} and \texttt{torch.cuda.max\_memory\_reserved} during the benchmark window.

\begin{table}[h]
\centering
\scriptsize
\setlength{\tabcolsep}{5.2pt}
\renewcommand{\arraystretch}{1.12}
\caption{\textbf{Runtime efficiency of FAITH/SeqFakeFormer (end-to-end decoding).}
Throughput and latency include full autoregressive decoding up to length 4 with early stopping.
AMP is enabled, batch size is 16, warm-up is 10 batches, and timing is over 512 images.}
\label{tab:faith_efficiency}
\begin{tabular}{c c c c c c}
\toprule
\textbf{Params} & \textbf{Batch} & \textbf{Throughput} & \textbf{Latency} & \textbf{Peak Mem} \\
\midrule
33.51 (M) & 16 & 100.15 (img/s) & 9.98 (ms/img) & 927.17 (MB) \\
\bottomrule
\end{tabular}
\end{table}

\noindent\textbf{Additional memory note.}
In the same run, the peak reserved memory is 1064.0 MB, while the current allocated memory after benchmarking is 137.21 MB.
We report both allocated and reserved memory in the released benchmark logs for completeness.

\subsection{Error-Mode Breakdown by Sequence Length}
To better understand failure cases under sequential provenance tracing, we categorize prediction errors into mutually exclusive types, computed per sample by comparing the non-padding predicted sequence $\hat{\mathbf{y}}$ and ground-truth $\mathbf{y}$:
\textbf{(i) Full-correct}, exact match in both order and length;
\textbf{(ii) Early-stop}, $\hat{T}<T$ (premature termination);
\textbf{(iii) Over-length}, $\hat{T}>T$ (spurious extra edits);
\textbf{(iv) Swap-only}, $\hat{T}=T$ and the multiset of predicted attributes matches the ground-truth but with a different order;
\textbf{(v) Wrong-attribute}, all remaining mismatches (including wrong attribute identity within the overlap).
This analysis directly complements Full-Acc by revealing whether failures are dominated by length errors, order errors, or attribute confusion.

Table~\ref{tab:faith_error_by_L} reports the breakdown by ground-truth sequence length $L\in\{1,2,3,4\}$.
We use a balanced evaluation with $N{=}2000$ samples per length to avoid dominance by any single length group.
As $L$ increases, the dominant error mode transitions from mild over-length or attribute mistakes to \emph{order errors} (Swap-only), indicating that recovering the correct temporal ordering becomes the primary bottleneck for long edit chains.
Notably, Swap-only accounts for 35.25\% of errors at $L=3$ and 67.85\% at $L=4$, consistent with the fact that later edits can partially overwrite earlier traces and make order-specific cues weak in the final image.

\begin{table}[t]
\centering
\small
\setlength{\tabcolsep}{4.6pt}
\renewcommand{\arraystretch}{1.12}
\caption{\textbf{Error-mode breakdown by ground-truth length $L$.}
Values are percentages and sum to 100\% within each $L$.}
\label{tab:faith_error_by_L}
\begin{tabular}{c c c c c c}
\toprule
\textbf{$L$} & \textbf{Full-correct} & \textbf{Early-stop} & \textbf{Over-length} & \textbf{Swap-only} & \textbf{Wrong-attribute} \\
\midrule
1 & 93.01 & 0.00 & 1.24 & 0.00 & 5.75 \\
2 & 77.10 & 1.93 & 2.53 & 10.41 & 8.03 \\
3 & 50.35 & 6.52 & 3.25 & 35.01 & 4.87 \\
4 & 24.74 & 6.63 & 0.00 & 67.85 & 0.78 \\
\bottomrule
\end{tabular}
\end{table}

\noindent\textbf{Implications for the three metrics.}
Order-dominant failures (Swap-only) sharply reduce Full-Acc because exact match requires the correct temporal ordering.
Fixed-Acc penalizes Swap-only errors at the mismatched positions, while Adaptive-Acc can remain relatively higher when early steps match but later order diverges.
Length errors (Early-stop and Over-length) are penalized by Fixed-Acc due to padding mismatches and by Full-Acc due to length mismatch, while Adaptive-Acc is comparatively less sensitive because it evaluates only the overlapping prefix.

\end{document}